\documentstyle[emulateapj,apjfonts]{article}

\def\mkfigbox#1#2{
\centerline{
\hbox{\epsfxsize=#1 \epsfbox{#2} \relax}
}
}

\def\Fig{{\footnotesize {\sc Fig.\/}~\,\thefigure. (continued) ---}}

\def\kms{km s$^{-1}$} 
\def\etal{{\it et al.}}

\def\Sec{${}^{\prime\prime}$\llap{.}}

\def\etal{{\it et~al.\/}}

\def\kms{{km~s$^{-1}$}}
\def\kpc-1{{kpc$^{-1}$}}
\def\Mpc-1{{Mpc$^{-1}$}}
\def\s-1{{sec$^{-1}$}}
\def\pdeg2{{deg$^{-2}$}}

\def\h0{{H$_0$}}
\def\q0{{$q_0$}}

\def\expector{E\/{\sc xpector\/}}
\def\expectorp{\expector\ }
\def\rms{{\it rms\/}}

\def\etal{{\it et al.\/}}
\def\kms{\hbox{$\rm km\,s^{-1}$}}

\def\ltsima{$\scriptscriptstyle \; \buildrel < \over \sim \;$}
\def\simlt{\lower.3ex\hbox{\ltsima}}

\def\gtsima{$\scriptscriptstyle \; \buildrel > \over \sim \;$}
\def\simgt{\lower.3ex\hbox{\gtsima}}

\def\about{\raise.3ex\hbox{$\scriptscriptstyle \sim $}}

\def\Sec{\hbox{${}^{\prime\prime}$\llap{.}}}

\def\sqr#1#2{{\vcenter{\vbox{\hrule height.#2pt
        \hbox{\vrule width.#2pt height#1pt \kern#1pt
        \vrule width.#2pt}
        \hrule height.#2pt}}}}
\def\square{{\mathchoice\sqr62\sqr62\sqr{4.2}1\sqr{3}1}\,}

\lefthead{Kelson \etal}

\slugcomment{Accepted 23 August 1999}

\begin{document}

\title{THE EVOLUTION OF EARLY-TYPE GALAXIES IN DISTANT CLUSTERS II.:
INTERNAL KINEMATICS OF 55 GALAXIES IN THE $z=0.33$ CLUSTER
CL1358+62$^1$} 

\author{Daniel D.  Kelson\altaffilmark{2,3}, Garth D.
Illingworth\altaffilmark{3}, Pieter G. van Dokkum\altaffilmark{4,5}, and
Marijn Franx\altaffilmark{5}}

\altaffiltext{1}{Based on observations obtained at the W. M. Keck
Observatory, which is operated jointly by the California Institute of
Technology and the University of California.}

\altaffiltext{2}{Department of Terrestrial Magnetism, Carnegie Institution
of Washington, 5241 Broad Branch Rd., NW, Washington, DC 20015}

\altaffiltext{3}{University of California Observatories / Lick Observatory,
Board of Studies in Astronomy and Astrophysics, University of California,
Santa Cruz, CA 95064}

\altaffiltext{4}{Kapteyn Astronomical Institute, P.O. Box 800, NL-9700 AV,
Groningen, The Netherlands}

\altaffiltext{5}{Leiden Observatory, P.O. Box 9513, NL-2300 RA, Leiden, The
Netherlands}

\begin{abstract}

We define a large sample of galaxies for use in a study of the
fundamental plane in the intermediate redshift cluster CL1358+62 at
$z=0.33$. We have analyzed high resolution spectra for 55 members of
the cluster. The data were acquired with the Low Resolution Imaging
Spectrograph on the Keck I 10m telescope. A new algorithm for
measuring velocity dispersions is presented and used to measure the
internal kinematics of the galaxies. This algorithm has been tested
against the Fourier Fitting method so the data presented here can be
compared with those measured previously in nearby galaxies.

We have measured central velocity dispersions suitable for use in a
fundamental plane analysis. The data have high $S/N$ and the resulting
random errors on the dispersions are very low, typically $<5\%$.
Uncertainties due to mismatch of the stellar templates has been
minimized through several tests and the total systematic error is of
order $\about 5\%$. Good seeing enabled us to measure velocity
dispersion profiles and rotation curves for most of the sample and
although a large fraction of the galaxies display a high level of
rotation, the gradients of the total second moment of the kinematics
are all very regular and similar to those in nearby galaxies. We
conclude that the data therefore can be reliably corrected for
aperture size in a manner consistent with nearby galaxy samples.

\end{abstract}

\keywords{ galaxies: evolution, galaxies: kinematics, galaxies: elliptical
and lenticular, cD, galaxies: structure of, galaxies: clusters: individual
(CL1358+62)}

%%%%%%%%%%%%%%%%%%%%%%%%%%%%%%%%%%%%%%%%%%%%%%%%%%%%%%%%%%%%%%%%%%%%%%%%

\section {Introduction}

Scaling relations have been extensively used to measure the evolution
of galaxies to redshifts of $z\sim 1$. However, without mass scales,
provided by measurements of rotation curves or line-widths, it is
impossible to extract the evolution of galactic $M/L$ ratios from that
of the underlying luminosity function. The explicit use of galaxy
masses, in the fundamental plane (\cite{faber87,dd87}), or the
Tully-Fisher relation (\cite{tf77}), makes these scaling relations
extremely powerful tools for measuring galaxy evolution.

The fundamental plane (FP) is useful for quantifying the luminosity
evolution of early-type galaxies in distant clusters. It is an
empirical relation between effective radius, velocity dispersion, and
surface brightness, for early-type galaxies. In a study of 10 nearby
clusters J\o{}rgensen {\it et al.\/} (1996) found
\begin{equation}
\log r_e \propto 1.24\log \sigma -0.82 \log \langle I\rangle_e,
\end{equation}
with an \rms\ scatter of $\pm 0.084$ in $\log r_e$ (a 21\% distance
uncertainty). The virial theorem and homology imply
\begin{equation}
\log {M\over L} \propto 2\log\sigma -\log r_e - \log \langle I\rangle_e
\end{equation}
The implication is that the mass-to-light ratio is a function of structural
and kinematical observables (\cite{faber87}), such that
\begin{equation}
M/L_V \propto r_e^{0.22}\sigma^{0.49}.
\end{equation}
The scatter in Coma of 23\% in $M/L_V$ (\cite{jfk93}), for a given $\sigma$
and $r_e$, suggests a small scatter in {\it age\/} for a given $\sigma$ and
$r_e$ (\cite{prugniel}). It has been postulated that systematic
departures from homology exist along the fundamental plane ({\it
e.g.\/}, \cite{ciotti}), but no satisfactory model has been created to
explain both the slope, and the low scatter of the scaling relation.
Evolution of the fundamental plane zero-point, slope, and scatter can
provide critical insight into the evolution of stellar populations of
early-type galaxies and may shed light on the origin of the
fundamental plane itself.

In a pioneering effort, Franx (1995) used fundamental plane
observations in A665 ($z=0.18$) to detect the evolution of stellar
populations in early-type galaxies. van Dokkum \& Franx (1996)
continued this technique with early-type galaxies in CL0024+16
($z=0.39$) and measured evolution consistent with the passive
evolution of stellar populations which formed at very high redshift.
More recently, Kelson \etal\ (1997) extended fundamental plane
measurements to small samples in CL1358+62 at $z=0.33$, and MS2053--04
at $z=0.58$. Their results confirmed the evolution in the fundamental
plane zero-point with redshift. van Dokkum \etal\ (1998b) have now
extended fundamental plane measurements to $z=0.83$, and have
constrained both the formation epoch of early-type galaxies and
$\Omega$. Other authors have also used the elliptical galaxy scaling
relations to constrain galaxy evolution and cosmology as well ({\it
e.g.\/}, \cite{schade,ellis97,ziegler,bender98,pahrethesis}), also
finding that E/S0s in clusters are uniformly old.

Both van Dokkum \& Franx (1996) and Kelson \etal\ (1997) suggested
that the FP tilt may have evolved between $z=0.33\hbox{-}0.58$ and the
present epoch, such that low-mass early-types are younger than their
high-mass counterparts. Their data also indicated that the scatter was
comparable to the nearby FP (\cite{jfk93,kelson}). However, their
samples were simply not large enough to accurately determine the slope
or scatter in the fundamental plane at intermediate redshifts. Kelson
\etal\ (1997) also found that the velocity dispersions of two E+A
galaxies were low ($\sigma \sim 100$ km/s). Are such low velocity
dispersions typical, and are these galaxies rotationally supported,
such as would be expected for disk-dominated systems ({\it e.g.\/},
\cite{franx93})?

In this paper we describe the data and reduction procedures used to
determine the internal kinematics of a large sample of galaxies in the
cluster CL1358+62. The HST data, from which the sample was selected,
are described in Kelson \etal\ (1999a) and van Dokkum \etal\ (1998a).
Extensive Keck spectroscopy has now enabled us to measure velocity
dispersions, dispersion profiles, and many absorption line rotation
curves, for a homogeneous sample of cluster members. In \S
\ref{selection}, we report on the sample selection for the
spectroscopy. The observations and data reduction procedures are
briefly discussed in \S \ref{data}. The stellar templates are prepared
in \S \ref{prep}. Our method for deriving velocity dispersions using a
direct fitting method is introduced in \S \ref{disp}. In \S
\ref{compare}, we test the results against other velocity dispersion
routines. In \S \ref{select}, we discuss the strategy for selecting
the best template and investigate the magnitude of template mismatch.
In \S \ref{final}, we discuss the aperture corrections required to
place our velocity dispersions on the same system as the J\o{}rgensen
\etal\ (1996) database, and use the spatially resolved kinematics to
compare the CL1358+62 members with nearby early-type galaxies. Lastly,
sources of error, both random and systematic, are discussed in \S
\ref{errors}.

%%%%%%%%%%%%%%%%%%%%%%%%%%%%%%%%%%%%%%%%%%%%%%%%%%%%%%%%%%%%%%%%%%%%%%%%

\section{Sample Selection}
\label{selection}

We are currently studying, in detail, the galaxy populations of three
clusters, CL1358+62 ($z=0.33$), MS2053--04 ($z=0.58$), and MS1054--03
($z=0.83$), selected from the Einstein Medium Sensitivity Survey
(\cite{gioia}). Determining membership is critical for studying any
cluster population. Fabricant \etal\ (1991) and Fisher \etal\ (1998)
have recently collected several hundred redshifts in a $10' \times
11'$ field around CL1358+62; the Fisher \etal\ (1998) sample is more
than 80\% complete at $R\le 21$ mag.

This field was targeted for extensive imaging with the HST WFPC2. A
two-color mosaic of the cluster CL1358+62 covering $64\, \square\, '$
was constructed using 12 pointings. Of the Fisher \etal\ (1998)
database, 194 spectroscopically confirmed cluster members fall within
the field of view of the HST imaging. From this large catalog, we
selected a sample for detailed study with the Low Resolution Imaging
Spectrograph (LRIS; \cite{okelris}) at the W.M. Keck Observatory. The
high resolution of the HST imaging allows us to derive structural
parameters with the accuracy needed for the fundamental plane
(\cite{kelson99a}).

From the catalog of cluster members within the HST mosaic, we randomly
selected galaxies down to $R < 21$ mag. As is discussed in more detail
in Kelson \etal\ (1999a), the $R$-band selected sample contains mostly
early-type systems, ranging from ellipticals to early-type spirals.
Our sample was constructed without regard for morphological
information, contrary to most low redshift samples. The selection was
performed with an effort to efficiently construct multi-slit plates
for LRIS. We used three masks, with different position angles on the
sky. During the designing of the multi-slit masks, we utilized the
apparent morphologies of several galaxies for major-axis positioning,
without impacting the selection criteria. The three mask designs
overlap in the core of the cluster, and so the sample is concentrated
towards the cluster center. A total of 67 slitlets were allocated for
the three masks.

\begin{figure}
\mkfigbox{5in}{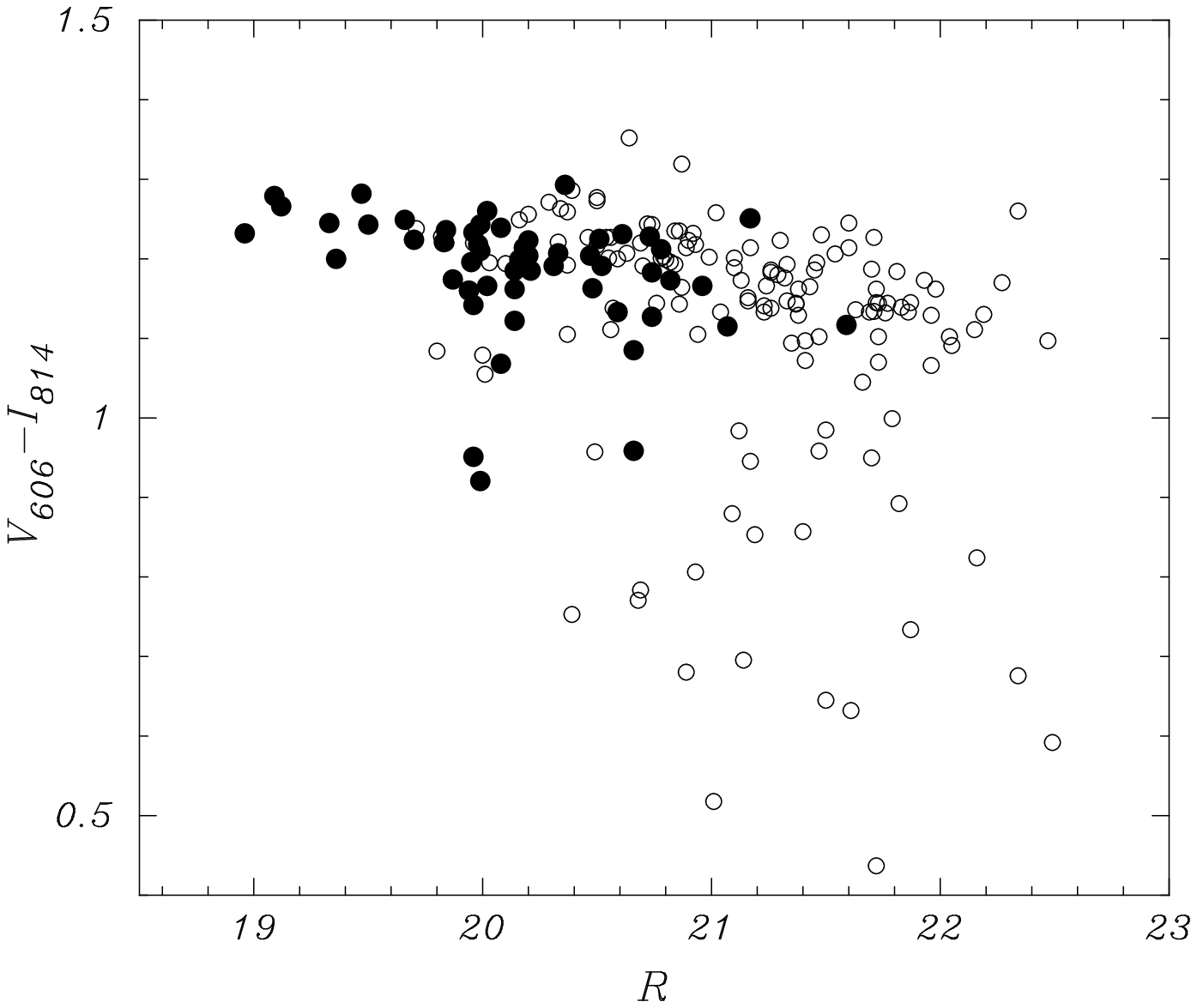}
\caption[fig1.ps]{
The color-magnitude diagram for the cluster in $R$ and $V_{606}-I_{814}$
(\cite{vdcm}). All 194 confirmed cluster members in the HST mosaic are
shown. The filled circles represent those galaxies in the fundamental plane
sample.
\label{cmd}}
\end{figure}

In Figure \ref{cmd}, we show the color-magnitude diagram for the
spectroscopically confirmed cluster members within the HST mosaic. At
the redshift of the cluster, $L^*$ corresponds to $R\sim 20.1$ mag
(adopting the $M/L_V$ evolution as measured by \cite{kelson}). There
is clearly a tight color-magnitude relation in the cluster, as shown
by van Dokkum \etal\ (1998a). The entire catalog of spectroscopically
confirmed cluster members within the HST mosaic is shown by the
circles, and those selected for fundamental plane analysis are shown
by the filled circles. As can be seen in the figure, the fundamental
plane sample fairly represents the cluster population, minus the very
bluest and faintest objects (\cite{vdcm}).

Fisher \etal\ (1998) found 108 members brighter than $R=21$ mag within
the HST mosaic. Our high-resolution spectroscopic sample contains 52
of them. Since three cluster galaxies fainter than the magnitude limit
were added, the total number of the sample presented here is 55. Two
of these galaxies were observed using more than one slit-mask. Eight
overlap the sample of Kelson \etal\ (1997).

This high-resolution sample of cluster members contains mostly
spectroscopically early-type galaxies, a few E+A galaxies, and a few
galaxies fainter than $R=21$ mag. The E+A galaxies are defined by $\rm
(H\delta+H\gamma+H\beta)/3 > 4$ \AA\ and [OII] 3727 \AA\ $< 5$ \AA\
(\cite{fish}). The E+A fraction in our high-resolution spectroscopic
sample is consistent with the fraction Fisher \etal\ (1998) found for
the cluster, about 5\%. By estimating the mass scales for these
galaxies, we hope to constrain the relationship between the
Butcher-Oemler effect ({\it e.g.\/}, \cite{bo78,dg83,cr97,cr98}) and
the histories of ``normal'' cluster members ({\it e.g.\/},
\cite{franx93,kelson}).

%%%%%%%%%%%%%%%%%%%%%%%%%%%%%%%%%%%%%%%%%%%%%%%%%%%%%%%%%%%%%%%%%%%%%%%%

\section{Observations and Data Reduction}
\label{data}

In May 1996 we obtained high $S/N$ two-dimensional spectra in the
field of CL1358+62. The data were acquired at the Keck I 10m telescope
on Mauna Kea using LRIS with the three multi-slit aperture plates. The
first two were exposed for 8100 s, and the third for 7400 s. We used
the 900 mm$^{-1}$ grating, with a dispersion of approximately 0.85
\AA\ per pixel. The restframe coverage was typically \about 4100 \AA\
to \about 5200 \AA. The seeing was typically 0\Sec 75 to 0\Sec 8
(FWHM). The slitlet widths were 1\Sec 05, resulting in a typical
resolution of $\sigma_i \approx 1.2$ \AA\ (\about 60 \kms). Some
slitlets were ``tilted'' with respect to the primary axis of the
slit-masks, to align them along galaxy major-axes. These slitlets have
broader projected widths, and leading to slightly poorer spectral
resolution. The treatment of the resolution is discussed explicitly in
\S \ref{prep}.

For the purposes of efficient and accurate extraction of one- and
two-dimensional spectra, a new set of FORTRAN procedures was created. These
subroutines are collected in a stand-alone package called \expector.
\expector's primary functions are that of flat-fielding, rectifying, and
wavelength calibrating two-dimensional multi-slit spectroscopic data.
\expectorp can be used trivially for long-slit reductions as well.

%%%%%%%%%%%%%%%%%%%%%%%%%%%%%%%%%%%%%%%%%%%%%%%%%%%%%%%%%%%%%%%%%%%%%%%%

\subsection{Preprocessing}

Some initial processing is required before execution of \expector. The
over-scan and bias-frame subtraction was done in IRAF. Cosmic-rays
were rejected using median filtering, with a search algorithm for
extended sharp objects. Care was taken to not degrade galaxy, or night
sky features in the removal of cosmic-rays. Next, the camera, or
``$y$''-distortion was determined by tracing the slit boundaries. This
distortion was removed from the flat-field images and the cosmic-ray
cleaned, bias-subtracted data frames.

The next steps involved flat-fielding the data, which requires several
stages. First, the spectral dome-flats were normalized, on a
slitlet-by-slitlet basis. The normalization occurs in both the $x$ and
the $y$ directions and produces two output images: a two-dimensional
pixel-to-pixel flat-field, and a one-dimensional column vector which
represents the (multi-slit) slit function. After the first pass of the
normalization procedures, re-normalized columns are medianed to
produce a column vector which is output as the slit function. The
pixel-to-pixel flat-field also contains a map of the fringing. For the
data presented here, however, fringing is negligible. This processing
of the flat-fields requires that the $y$-distortion be removed
beforehand, so that the individual two-dimensional spectra are aligned
with the dispersion axis of the CCD.

The data frames are divided by the pixel-to-pixel flat-fields. Tests
using spectra extending to $\sim 9500$ \AA\ showed that this step can
be effective in removing fringing. Next, the one-dimensional slit
functions are shifted (in the $y$ direction), such that the slit edges
match those in the data frames. These shifts are required if the slit
edges in the CCD images of the flat-fields are not registered with the
slit edges in the data frames. Finally, the columns in the data frames
are divided by the shifted slit functions. Each mask has its own set
of flat-fields, in order to best match the spectral coverage of each
pixel, and to match the individual slit functions.

%%%%%%%%%%%%%%%%%%%%%%%%%%%%%%%%%%%%%%%%%%%%%%%%%%%%%%%%%%%%%%%%%%%%%%%%

\subsection{Rectification and Wavelength Calibration}

The two-dimensional multi-slit images were rectified and wavelength
calibrated by \expector\ using the night sky lines ([O {\sc I}], Na
{\sc I}, and OH). A single $2048 \times 2048$ multi-slit image can be
fully rectified, wavelength calibrated, and rebinned, for example, in
just a few minutes on a DEC Alpha processor. We summarize the
procedures below. The interested reader can find more details in
Kelson (1998).

The technique for rectifying two-dimensional spectra uses Fast Fourier
Transforms and efficient cross-correlations to measure the pixel
shifts between regions of sky (or lamp) lines from image row to image
row. Thus, our method does not rely on centering algorithms to fix the
positions of {\it individual\/} lamp, or night sky, emission lines.
The shifts, in pixels, between adjacent CCD rows are measured as a
function of wavelength, by performing the cross-correlations in
subsections of each spectrum, divided along the dispersion direction.
We fit a two-dimensional polynomial to the shifts as a function of
$(x,y)$ position, and use this polynomial to align all the rows to a
common coordinate (wavelength) system. The typical \rms\ scatter about
the fit for these rectification transformations is a few millipixels,
with several tens of millipixels in the worst (tilted) cases. This
method for rectifying two-dimensional spectra is extremely accurate
and computationally economical due to its maximal use of the available
data.

Because the rectification aligns all the CCD rows, within a given
spectrum, one only needs to determine the wavelength solution for a
single row (or average of several, in order to improve the $S/N$).
\expector's automated wavelength calibration routine can use LRIS lamp
lines, or the night sky emission lines. The algorithm quickly finds
the emission peaks, measures the FWHM of each peak non-parametrically,
and defines the center of each line to be the average of the two
half-power points. When calibrating spectra using night-sky emission
lines, the algorithm quickly finds an approximate central wavelength
using cross-correlations, and then identifies triplets of P1 and P2 OH
lines using ratios of spacings between sky lines. After an initial
determination of a low-order wavelength solution, more lines are
identified and used to refine the solution.

We chose to fit a fifth order Legendre polynomial for the dispersion
solutions. The \rms\ scatter about the fit was typically $\pm
0.01$-$0.03$ \AA, though somewhat worse for tilted slitlets due to the
increase in the width of the lines. Comparison lamp spectra were used
to test the validity of the wavelength solution as derived from the
night sky emission. The sky lines had equivalently low \rms\ scatter
about the lamp solutions, once the zero-point in the wavelength
calibration had been corrected. We opted for a dispersion solution
derived directly from the night sky, using emission lines down to 5224
\AA.

Each slitlet was rebinned logarithmically to a common wavelength
range, by chaining the rectification transformations and dispersion
solutions together. This step involves a single interpolation in the
$x$ (wavelength) direction. The output images are extra wide to
accommodate the staggering of the slitlets in the dispersion
direction, and flux was conserved in the rebinning of each image.
Note: the data were rebinned only once in the $y$ direction of the CCD
and once in the $x$ direction. The exposures were then averaged,
weighting inversely by the expected noise, including the noise in the
sky and galaxies, as well as the read noise in the amplifier.
Background subtraction was carried out in a manner similar to
long-slit reductions, in which the sky spectrum at the location of a
galaxy is interpolated using the observed sky on either side.
Residuals around strong night sky lines, such as 5577 \AA, and 6300
\AA, were interpolated over, though this step was performed largely
for cosmetic reasons.

\begin{figure}
\mkfigbox{5in}{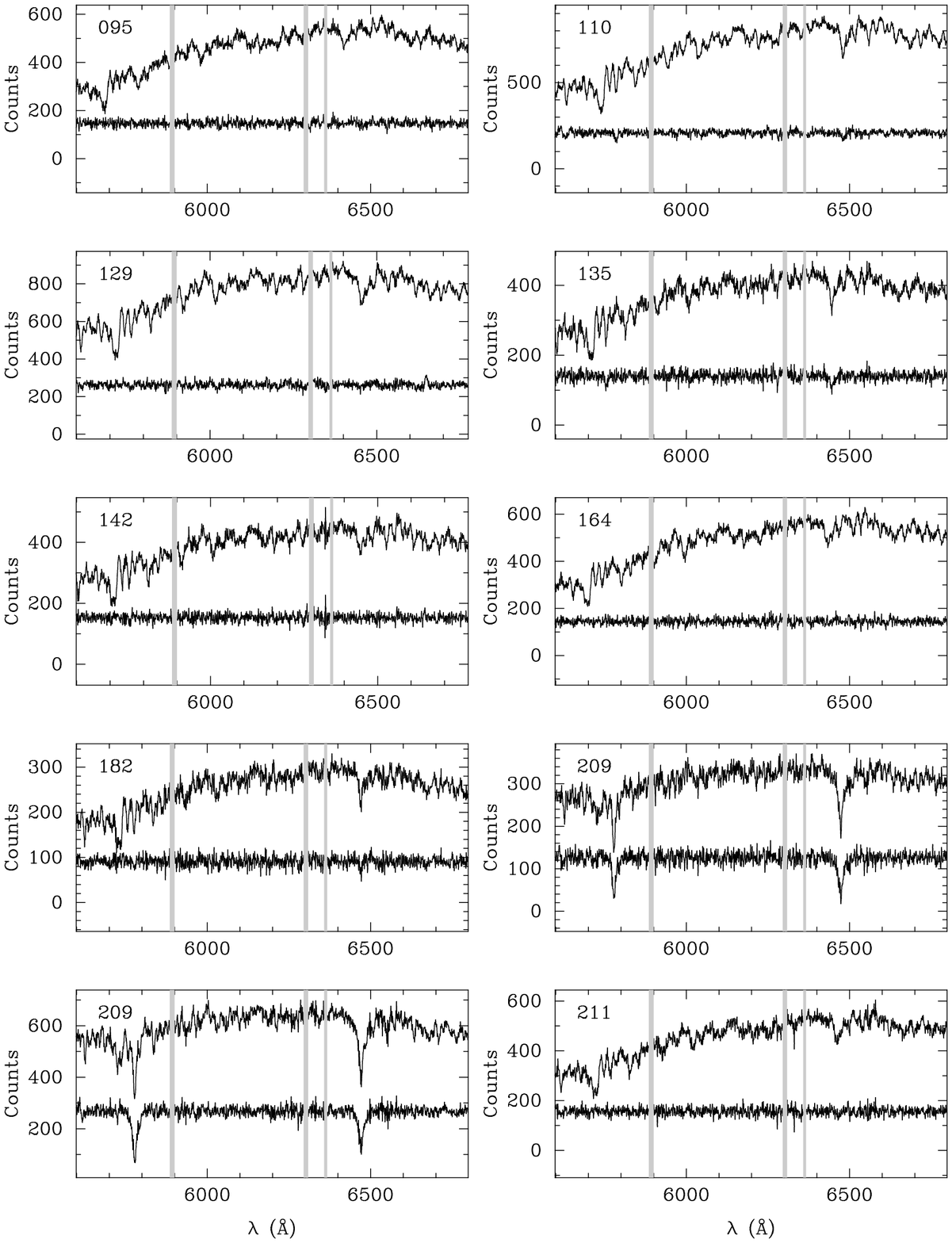}
\caption[fig2a.ps,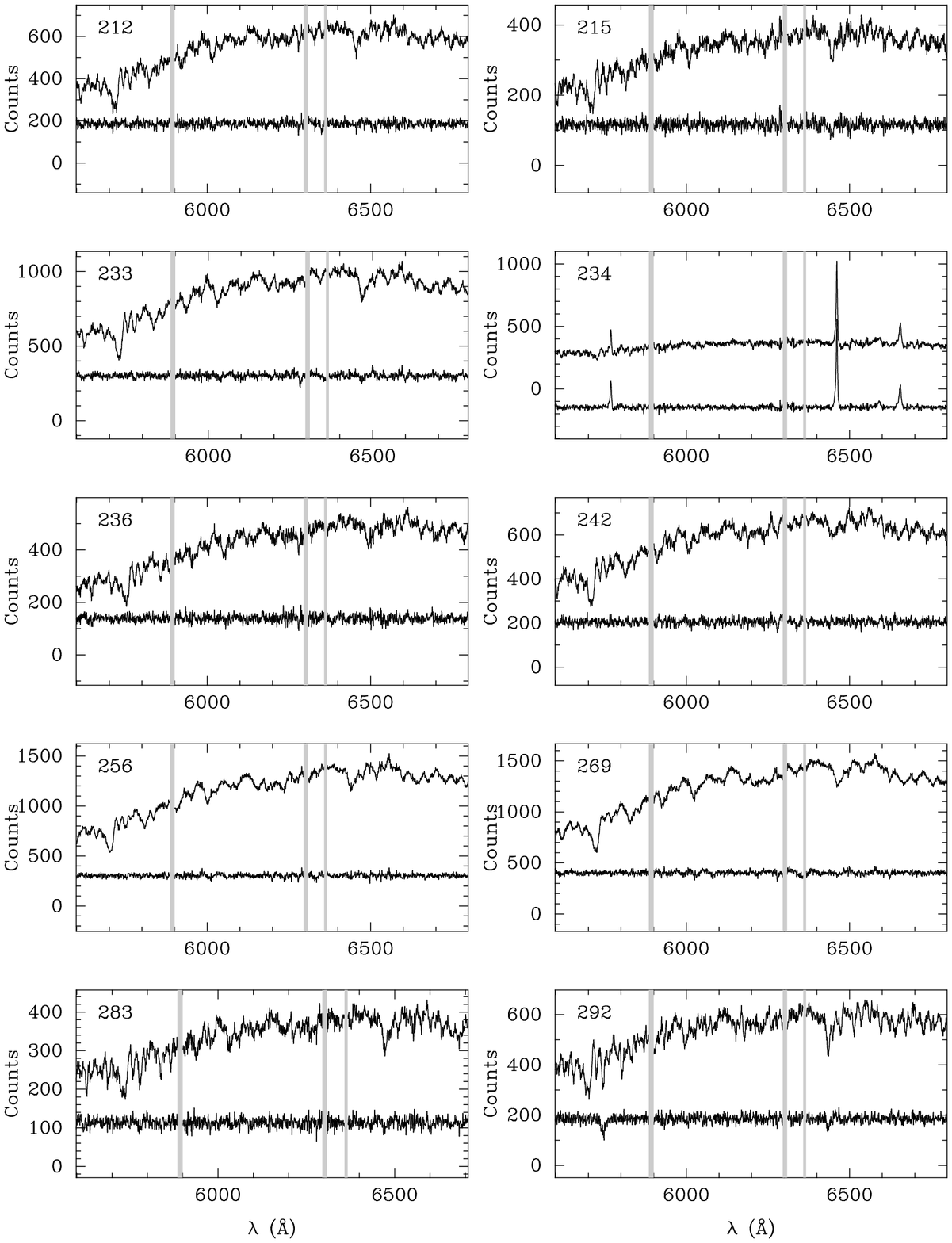,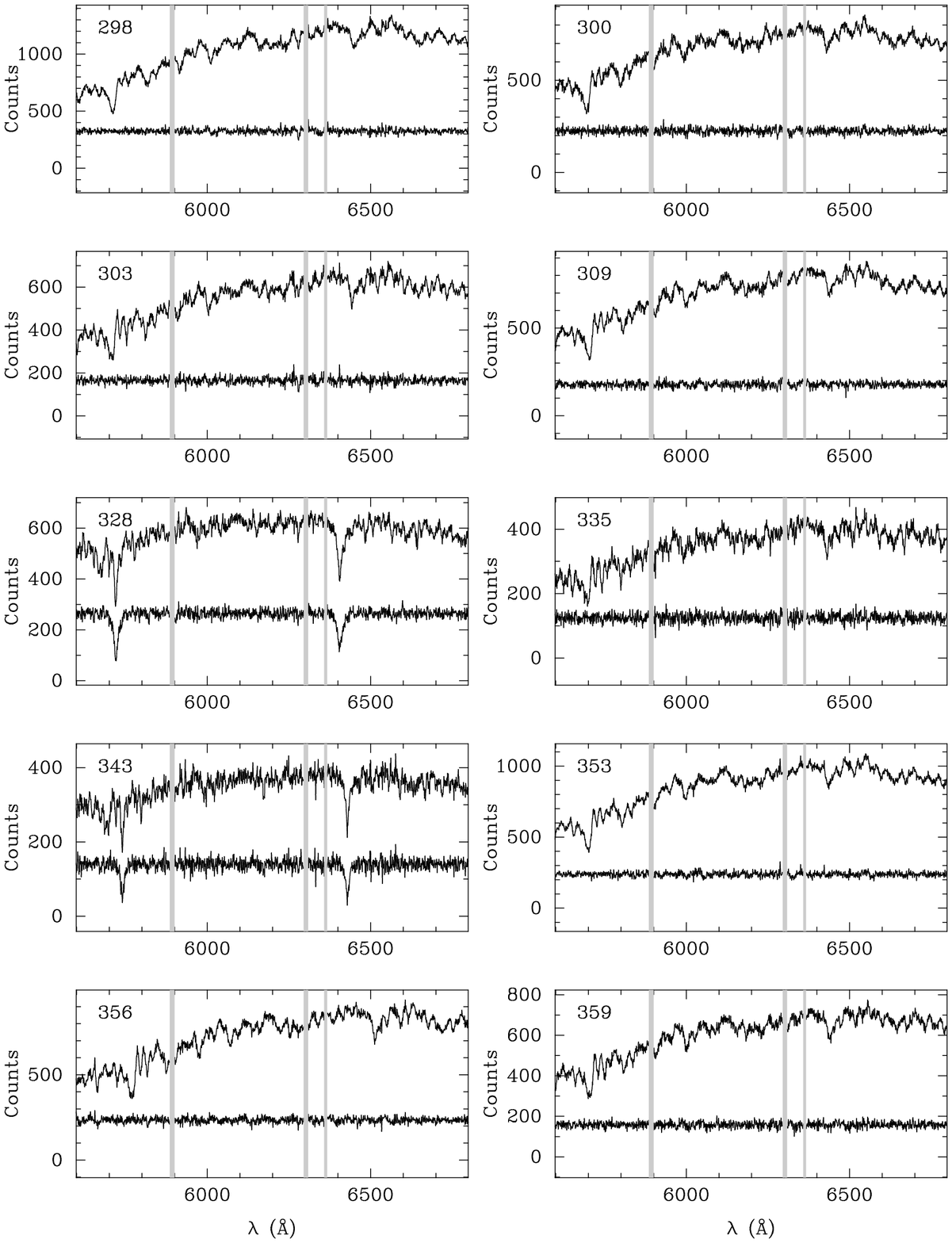,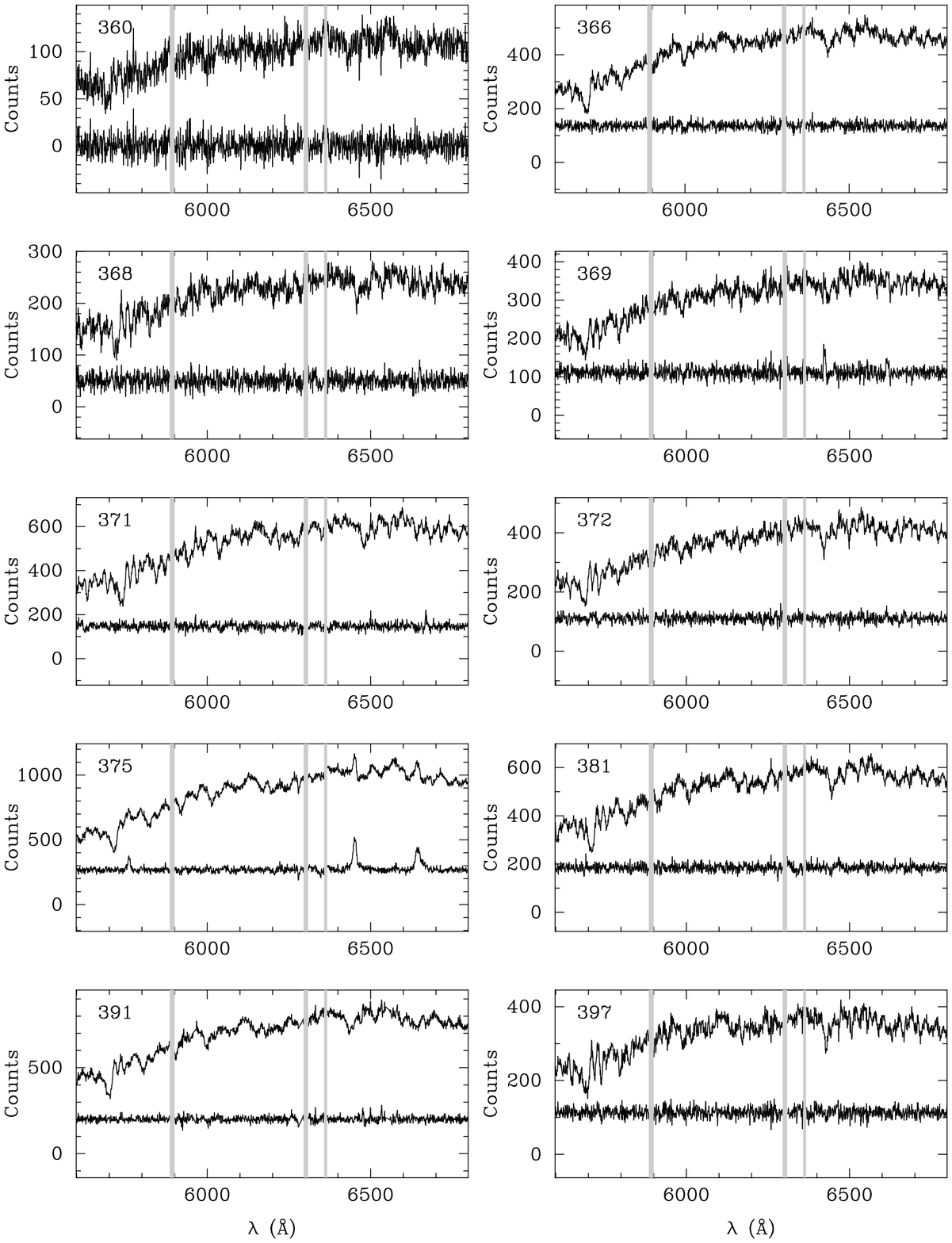,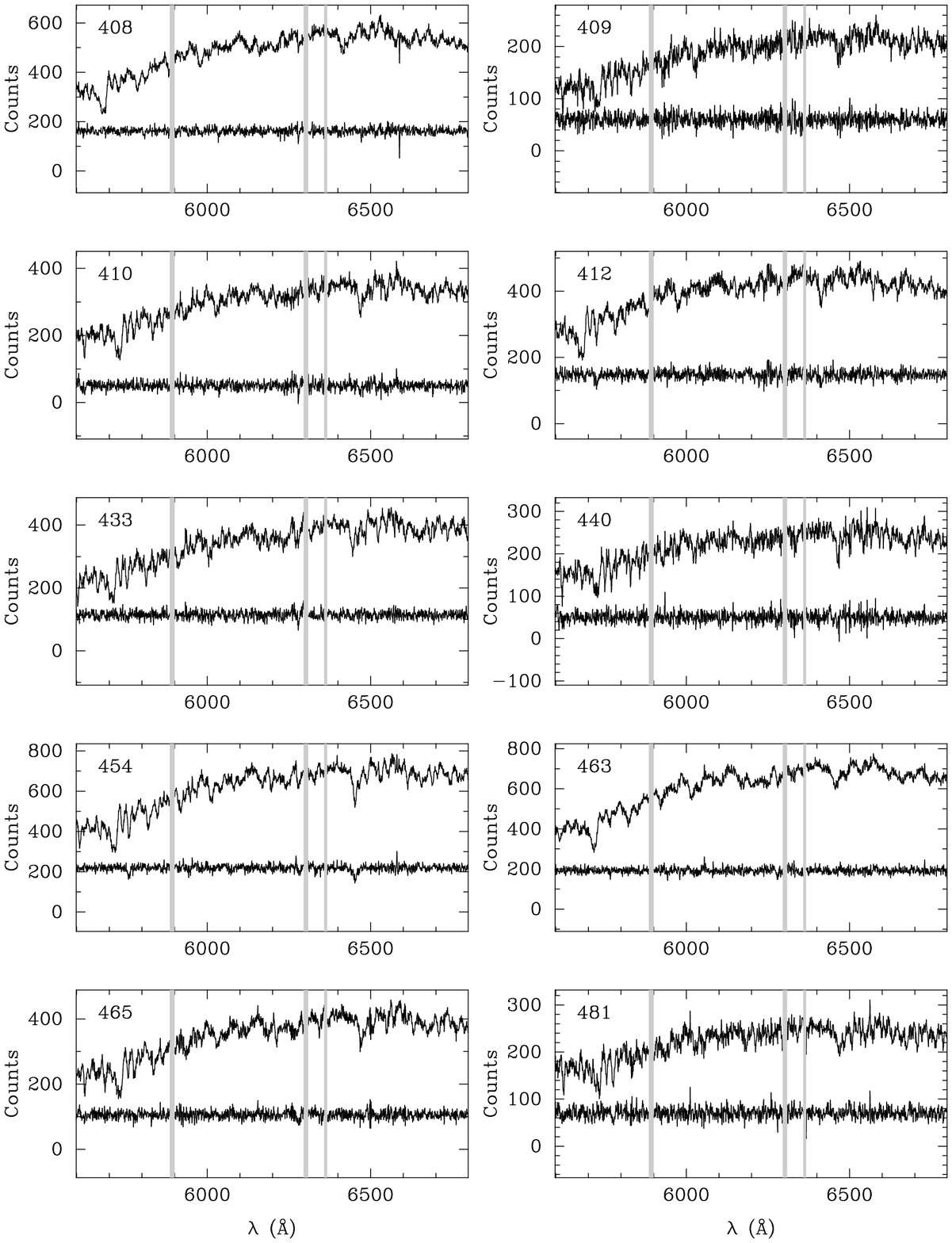,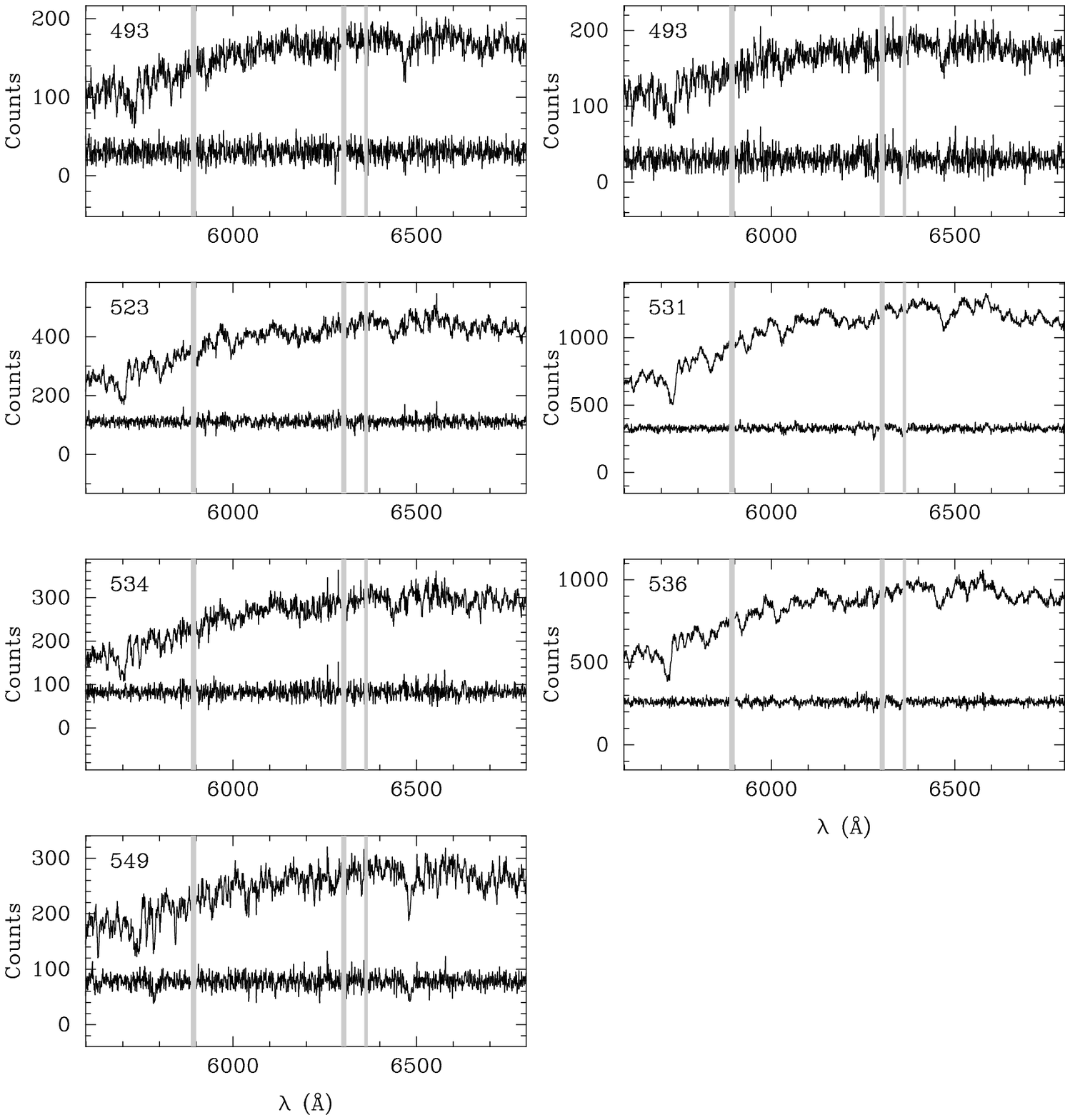]{
The galaxy spectra, extracted from an aperture of 5 CCD rows. These spectra
are {\it not\/} smoothed. Beneath each spectrum are the residuals from the
velocity dispersion fit using template HD72324. The regions of bright sky
emission lines around Na I, 6300 [O I] and 6364 [O I] are shown. Note the
wide range of Balmer line residuals. Signal-to-Noise estimates are
listed in Table \ref{sigfinal}
\label{spectra}}
\end{figure}

Spectra for the galaxies were extracted by summing five CCD rows. The
mean $S/N$ ranges from 10 to 80 per \AA, with a median of \about 35
per \AA. This high $S/N$ allowed us to sum the image rows directly,
rather than optimally, which would complicate the determination of the
effective aperture of the extraction. The spectra themselves are shown
in Figure \ref{spectra}. The 5 row aperture, $1\Sec 08 \times 1\Sec
05$, is equivalent to a circular aperture of diameter 1\Sec 23
(\cite{jfk95}). In an $\Omega=0.5$, $H_0=65\ \kms\rm\,Mpc^{-1}$
universe, this corresponds to a circular aperture of diameter \about
5.5 kpc (or equivalently $11''$ at the distance of Coma). Slitlets
tilted by $45^\circ$ have an effective aperture 19\% larger in
diameter.

%%%%%%%%%%%%%%%%%%%%%%%%%%%%%%%%%%%%%%%%%%%%%%%%%%%%%%%%%%%%%%%%%%%%%%%%

\section{Preparation of Template Spectra}
\label{prep}

Because these galaxies are at cosmological distances, one must process
stellar template spectra with care in order to ensure that they have
the the same {\it restframe\/} instrumental resolution as the galaxy
spectra (\cite{franx93}). For nearby galaxies, this is obviously not a
problem, as the same instrumental setup can be used to obtain both the
stellar and galactic spectra. However, for high-$z$ objects, template
spectra cannot be observed with the same instrumental setup and one
must carefully model the resolution of the galaxy spectra. The stellar
spectra must be treated with extra caution as well if their resolution
is similar to that in the galaxy spectra (see \cite{franx93,vdf96}).
The spectra of template stars must be convolved by a kernel to match
the point-spread function of the galaxy spectra.

Using LRIS during twilight in May and August 1996, we performed
long-slit spectroscopy of several late-type stars (G and K stars),
listed in Table \ref{templates}. The 1200 mm$^{-1}$ grating was used,
providing spectral coverage from 4000 \AA\ to 5300 \AA, with a
dispersion of approximately 0.64 \AA\ per pixel. These resulted in
spectra with higher resolution that that of our galaxy spectra (see \S
\ref{galres}). The telescope was moved in a raster pattern to expose
the star across the full width of the slit at several spatial
positions. It is important to fill the width of the slit (0\Sec 7),
such that the spectral resolution is limited by the slit width, and
not by the seeing. These data were also bias-subtracted, flat-fielded,
rectified, and wavelength-calibrated. The extracted stellar spectra
have very high $S/N$, several hundred per pixel.

%%%%%%%%%%%%%%%%%%%%%%%%%%%%%%%%%%%%%%%%%%%%%%%%%%%%%%%%%%%%

\subsection{The Wavelength Calibration of the Stellar Spectra}

The stellar spectra were wavelength calibrated using a sparse sample
of comparison lamp lines in the blue (mostly Ar and Hg). A dearth of
bright lamp lines in the blue led us to test the validity of the
dispersion solution when applied to the stellar long-slit spectra. We
performed a cross-correlation of a high resolution solar spectrum,
with the twilight and template spectra, over small, 100 \AA\ bins, to
provide an independent check on the wavelength calibration. There was
a high-order systematic trend with wavelength, such that there was a
net velocity difference between the red and blue sides of the stellar
spectra. The wavelength calibration derived from the internal lamps
was in error, by about $30$ \kms\ peak-to-peak, from blue to red, in
the May stellar spectra, and about $40$ \kms\ peak-to-peak in the
August stellar spectra --- equivalent to a non-linear stretching of
about 1 pixel. This can be attributed to the dearth of available lamp
lines in the blue, coupled with centering errors in asymmetric arc
profiles, especially in the August lamp data. Using this
cross-correlation information, we rebinned the raw template spectra to
the proper dispersion solution.

%%%%%%%%%%%%%%%%%%%%%%%%%%%%%%%%%%%%%%%%%%%%%%%%%%%%%%%%%%%%

\subsection{Spectral Resolution of the Template Stars}

\begin{figure}
\mkfigbox{3in}{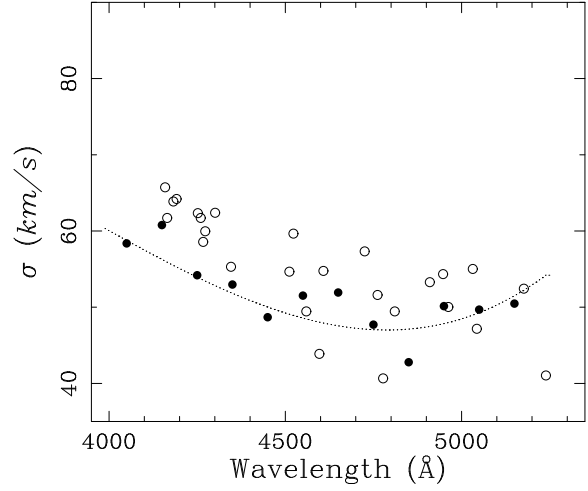}
\caption[fig3.ps]{
The resolution of the May 1996 stellar templates with wavelength, in the
frame of CL1358+62. The open circles denote the arc line widths, and the
filled circles show the resolution as derived using the solar
spectrum. The dotted line shows the fit of a cubic to the filled
circles. The resolution determined from the arc lines is similar to
that derived using the solar spectrum, and the match is especially
good over the range within which the velocity dispersion fitting will
be performed, from \about 4200 \AA\ to \about 5100 \AA, in the
restframe of the galaxies.
\label{arc}}
\end{figure}

We determined the instrumental broadening in our stellar spectra in
two complementary ways. Initially, the resolution of these stellar
spectra was determined using the widths of the comparison lamp lines.
For the May stellar spectra, the $\sigma$ widths of these lines are
shown in Figure \ref{arc} as crosses. However, the internal lamps in
LRIS do not necessarily illuminate the spectrograph in the same manner
as the sky. Thus, we wished to compare the arc line widths with the
resolution of the stellar spectra themselves. We were able to measure
the intrinsic broadening in the stellar spectra by comparing them with
a high-resolution solar spectrum.

Using the Fourier Fitting Method (\cite{fih89}), we determined the
broadening required to match the high-resolution solar spectrum to the
spectrum of the twilight sky. The typical resolution in the May 1996
stellar spectra was \about 50 km/s. These Fourier Fitting Method
results are shown as open circles in Figure \ref{arc}, for direct
comparison with the $\sigma$ widths of the lamp lines. The two methods
give similar broadening profiles as a function of wavelength. This
comparison gives us confidence that the resolution is well-determined.
We also used the high-resolution solar spectrum to find the
instrumental broadening in the stellar spectra themselves and found
consistent results, except around the Balmer lines, where the solar
spectrum is clearly not a good match to the later-type stars.

The August 1996 stellar template spectra were taken after an upgrade
of LRIS, in which a new CCD, field-flattener, and dewar were
installed. The lamp spectra taken at this time have strongly
asymmetric profile shapes, probably due to the internal illumination
overfilling the grating (\cite{gill}). Thus, the spectrograph's
resolution of the sky (and galaxies) may not be well-described by the
arc line profiles. Therefore, we used the high resolution solar
spectrum again, to directly measure the instrumental broadening in the
August stellar spectra. There was no twilight spectrum from August,
but the spectrum of the G5IV template star is close in type to the
solar spectrum, thus minimizing any problems with template mismatch in
the determination of the resolution. At 4500 \AA, the high-resolution
spectrum implies an instrumental broadening in the HD5268 spectrum of
45 \kms, while the lamp line widths yield 55 \kms.

If the resolution derived from the August 1996 arc spectra were to
have been adopted instead of that derived from the solar spectrum
comparison, we would have incurred a $-3\%$ error in velocity
dispersion at $\sigma=100$ \kms. At $\sigma=300$ \kms, this error is
only $-0.3\%$. Thus, the effect on the slope, $\alpha$, of the
fundamental plane, $\log{r_e}\propto \alpha \log\sigma+\beta
\log\langle I\rangle$, would have been to increase the coefficient of
$\log \sigma$, by $\Delta \alpha \approx +0.02$.

For both the May and August stellar spectra, we adopted the resolution
determined from the high resolution solar spectrum.

%%%%%%%%%%%%%%%%%%%%%%%%%%%%%%%%%%%%%%%%%%%%%%%%%%%%%%%%%%%%%%%%%%%%%%%%

\subsection{The Instrumental Resolution of the Galaxies}
\label{galres}

The night sky lines enabled us to measure the variation in
instrumental broadening as a function of wavelength, for each of the
CL1358+62 slitlets. However, care must be taken since blended sky
lines can bias the measurement of the instrumental broadening.
Therefore, the spectrograph resolution was measured using a list of
bright night-sky lines, chosen such that they are unblended at our
resolution. In particular, we used the emission lines at 5577 \AA,
6300 \AA, 6364 \AA, and an array of OH P1 lines, all of which are
listed in Table \ref{linetable} (see \cite{oster}). The OH P1 lines
are close blended doublets, which, for our instrumental setup, impacts
our estimate of the resolution at levels smaller than $< 0.5\%$. The
effect is negligibly small for our velocity dispersion measurements.
We also note that the profiles of these sky lines were well fit by
pure Gaussians.

The sky line widths vary as a function of wavelength and position
across the CCD. We fit third order polynomials to the sky-line widths
as a function of wavelength, for each slitlet. The variation in
template resolution with wavelength was also modeled with a third
order polynomial. The final, prepared template spectra were generated
by convolving the stellar spectra using a Gaussian with a $\sigma$
width determined by the quadrature difference between the two
resolution polynomials. This procedure was tested by measuring the
effective resolution of the prepared template spectra using the
high-resolution solar spectrum. The effective resolution of the
prepared templates agreed with the sky line widths to within a few
percent.

\begin{figure}
\mkfigbox{5in}{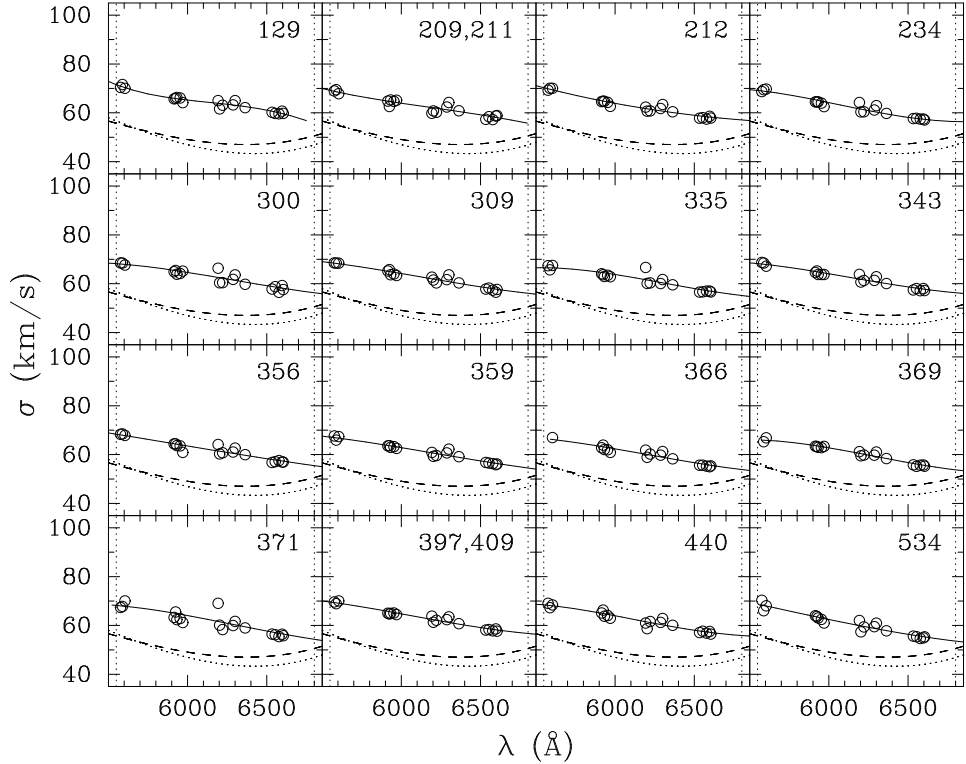}
\caption[fig4.ps]{
The instrumental resolution of the spectrograph as functions of wavelength
for the slitlets of several galaxies in one of the multi-slit masks. The
points represent the individual sky line widths for the lines shown in
Table \ref{linetable}. The solid line is the low-order polynomial fit to
these line widths. The dashed and dotted lines represent the resolution of
the stellar templates from May and August, respectively. The vertical
dotted lines bracket the wavelength range in which the galaxy kinematics
will be derived.
\label{resolution}}
\end{figure}

Figure \ref{resolution} shows sky line widths as a function of
wavelength for each of the CL1358+62 galaxy slitlets in the second
mask. The circles show the measured widths of the sky lines in \kms,
while the solid lines show the polynomial fit to these data. The
dashed and dotted lines show the polynomial representation of the
template spectral resolution for the May and August stellar spectra
(see, {\it e.g.\/}, Figure \ref{arc}). The resolution for each slitlet
was determined from the combined spectra of three exposures per mask.
The widths of the sky lines were also measured in the three individual
exposures separately to test the stability of the resolution. These
individual measurements agree with the data in Figure \ref{resolution}
to a high degree of accuracy ($\pm1\%$) and are therefore not shown.

Each slitlet was given its own suite of prepared template spectra,
broadened to the have the same restframe resolution as that of the
galaxy (or galaxies) contained within. In all cases, the resolution of
the template spectra was higher than that for the galaxies.

%%%%%%%%%%%%%%%%%%%%%%%%%%%%%%%%%%%%%%%%%%%%%%%%%%%%%%%%%%%%%%%%%%%%%%%%

\section{Velocity Dispersion Fitting Methods}
\label{disp}

We are endeavoring to find the line-of-sight velocity distribution
(LOSVD) which gives the best match between our galaxy spectrum and a
stellar template broadened by the LOSVD ({\it e.g.\/}, \cite{rix92}).
We have parameterized the LOSVD by a Gaussian, as has been done for
comparable studies of nearby samples of early-type galaxies. While
sometimes the line profiles of real galaxies can be significantly
non-Gaussian (see, {\it e.g.\/}, \cite{franx88}), deviations from
Gaussian LOSVDs are more typically of the order of 10\%
(\cite{bender94}), and are unimportant for studies of the fundamental
plane. We use two methods to determine the velocity dispersions, the
widely used Fourier Fitting method and a direct fitting method which
is more appropriate for high redshift galaxies (see below). We use
both algorithms in order to ensure that our measurements are
consistent with those in the nearby samples.

%%%%%%%%%%%%%%%%%%%%%%%%%%%%%%%%%%%%%%%%%%%%%%%%%%%%%%%%%%%%

\subsection{Fourier Fitting Method}

The parameterized LOSVD, has two parameters, the radial velocity $V$,
and the dispersion, $\sigma$, in this velocity. We therefore build a
model, $M$, of each galaxy spectrum, $G$, by a convolution $M =
B(V,\sigma) \circ T$, where $T$ is the template stellar spectrums and
$B$ is the broadening function. We can then compute the residuals
between the convolved model and the galaxy spectrum:
\begin{equation}
\chi^2=|G-M|^2
\label{eq:genfit}
\end{equation}
In the Fourier Fitting method (\cite{fih89}), this $\chi^2$ calculation
is performed in the Fourier domain, such that the Fourier transform
of the convolved template is matched to the transform of the galaxy
spectrum:
\begin{equation}
\chi^2=|\widetilde G-\widetilde M|^2
\end{equation}

When measuring velocity dispersions, we are most interested in finding
the broadening function which produces the best match to the
absorption features. While single stellar spectra can provide good
matches to galaxy absorption features, the continuum of a stellar
spectrum does not match the continuum of a galaxy spectrum well at
all. For our high redshift spectra, the detector response also alters
the continuum shape with respect to the templates. The low wavenumbers
which describe the general shape of galaxy and stellar spectra have
little, if anything, to do with the computational problem of deriving
LOSVDs, but can contribute substantially to the $\chi^2$ summation.

Therefore, an important aspect of measuring velocity dispersions is
the matching of the template continuum to that of the galaxy. In the
Fourier Fitting method (\cite{fih89}), low wavenumbers are filtered
during the summation of $\chi^2$. This filtering helps minimize
mismatch between stellar templates and galaxy spectra by essentially
using sines and cosines as additive functions to match the continuum
of the template to that of the galaxy:
\begin{equation}
\widetilde W = \cases {0, &$f_{\rm low} < f$; \cr
[1-\cos\bigl({ \pi {f-f_{\rm low}\over f_{\rm low}}}\bigr)]/2, 
& $f_{\rm low}\le f < 2f_{\rm low}$;\cr
1 & $2 f_{\rm low}\le f$.\cr}
\end{equation}
such that the $\chi^2$ summation can now be written as
\begin{equation}
\chi^2=|(\widetilde G-\widetilde M)\times \widetilde W|^2
\end{equation}
The weight vector equals zero, at frequencies at or below $f_{\rm
low}$, and unity, at frequencies at or greater than $2 f_{\rm low}$.
The filter is a continuous function of wavenumber so that the method
is insensitive to the precise cutoff, and can be thought of as having
an effective frequency cut of $f=1\case{1}{2} f_{\rm low}$. Put
another way, continuum matching by the Fourier Fitting Method consists
of specifically removing long-wave sines and cosines from both star
and galaxy. In order for our velocity dispersions to be comparable to
those of J\o{}rgensen \etal\ (1995), we adopt the same restframe
filter window of $k\approx [100 {\rm \AA}]^{-1}$ and use $f_{\rm
low}=9$.

%%%%%%%%%%%%%%%%%%%%%%%%%%%%%%%%%%%%%%%%%%%%%%%%%%%%%%%%%%%%

\subsection{Direct Fitting Methods}
\label{direct}

When measuring velocity dispersions of high redshift galaxies
additional, non-uniform sources of noise in the data can complicate
the analysis. At high redshift, galaxy spectra are redshifted to
wavelengths where bright night-sky emission lines are present. The
shot-noise in these lines are a strong source of local noise in the
resulting galaxy spectra. Furthermore, sharp residuals from the
subtraction of the flux in these lines adds high-frequency noise to
the extracted galaxy spectra. Therefore, we constructed a direct
fitting algorithm, which allows for a more complicated treatment of
the data.

%%%%%%%%%%%%%%%%%%%%%%%%%%%%%%%%%%%%%%%%%%%%%%%%%%%%%%%%%%%%

\subsubsection{Non-uniform Weighting}

In Fourier methods every pixel receives identical weight in the fit of
the broadened template to the galaxy spectrum. Uniform weighting,
however, is not always desired, especially when there exist highly
localized sources of noise in the data. Such noise adds high frequency
power to the power spectrum of the data, leading to additional
uncertainty in the $\chi^2$ fit. At best, one can attempt to globally
reduce the effects of such noise in Fourier methods using, {\it
e.g.\/}, Wiener filtering (\cite{simkin,bender90,rix92,jesus}).

By performing the $\chi^2$ computation in the real domain, however,
one can assign weights to individual pixels. Complicated weighting
procedures can be adopted, such as, for example, masking of regions of
poor sky-subtraction, weighting by the inverse of the expected noise,
or masking of specific spectral features such as extraordinarily
strong Balmer absorption lines or absorption features which have been
filled in by emission. This process allows such features to be
excluded from the $\chi^2$ calculation, leaving the results less
sensitive to problematic features and noisy regions of the data. In
performing a weighted least-squares fit to galaxy spectra, we modify
Equation \ref{eq:genfit} to include a vector of weights, $W$.
\begin{equation}
\chi^2=|(G-M)\times W|^2
\label{eq:wtfit}
\end{equation}

In order to properly treat the non-uniform sources of noise in our
CL1358+62 data, we developed a variant of the Rix \& White (1992)
algorithm which determines line-of-sight velocity distributions
(LOSVDs) by computing $\chi^2$ in the real domain. As discussed
earlier, we adopt a Gaussian function for the parameterized LOSVD.

When using either the Fourier Fitting method, or a direct-fitting
algorithm, one performs a least-squares fit of broadened templates to
the observed galaxy spectra. The only essential difference between the
two methods as applied here is the ability to use non-uniform
weighting of the data in the real $\chi^2$ summation. Thus, the
results of a fit in the pixel (real) domain with uniform weighting
should be identical to those found using Fourier fitting. This is
shown to be the case explicitly in \S \ref{compare}.

%%%%%%%%%%%%%%%%%%%%%%%%%%%%%%%%%%%%%%%%%%%%%%%%%%%%%%%%%%%%

\subsubsection{Continuum Matching}

As with the Fourier Fitting Method, matching the galaxy continuum is
an important component of our direct fitting method. For galaxies at
higher redshift, there is a multiplicative source of mismatch between
the galaxy and stellar continuum as well as the standard, additive
one.

When deriving velocity dispersions of nearby galaxies, one can obtain
template spectra using an identical instrumental setup. In this case,
any mismatch between the observed shapes of the galaxy and stellar
spectra arises because the overall shapes of galaxy spectra are not
perfectly represented using a single stellar spectral type; galaxy
spectra are complicated composites of many individual stellar spectra.
Such differences in the continuum can be approximated by adding a
function of wavelength comprised of simple functions, such as
polynomials (\cite{rix92}), to make up differences between the flux in
the template and the flux in the galaxy.

The second type of mismatch between the shapes of galaxy and stellar
spectra occurs because the spectra of distant galaxies are obtained
with a completely different instrumental setup from the stellar
template spectra. In this case, the galaxy and stellar spectra may
have been obtained using different gratings and detectors. Thus, the
spectra of the stars and the distant galaxies may have different
overall shapes because of the instrumental response function, which is
multiplicative.

Therefore, we incorporate a multiplicative polynomial, $P_M$, into the
fit to remove large-scale shape differences between the observed
stellar and galactic spectra which are largely due to changes in
instrumental setup, {\it i.e.\/}, differences in instrumental
throughput as a function of wavelength. The inclusion of $P_M$ in the
least-squares solution ensures that our results are insensitive to the
normalization or flux calibration of galaxy and stellar template
spectra (see below).

The incorporation of a multiplicative polynomial in the fit is a key
difference between previous procedures ({\it e.g.\/}, \cite{rix92})
and this particular direct-fitting algorithm. Instead of normalizing
the stellar template and galaxy spectra before performing the fitting
of the template to the galaxy, the normalization is explicitly
incorporated into the continuum matching process. Since this
normalization becomes a part of the $\chi^2$ minimization, the effects
of the continuum normalization on the derived velocity dispersions
are, by definition, minimized. The zeroth order term in this
multiplicative polynomial is equivalent to the ``line strength''
parameter, $\gamma$, in more traditional velocity dispersion
algorithms.

%%%%%%%%%%%%%%%%%%%%%%%%%%%%%%%%%%%%%%%%%%%%%%%%%%%%%%%%%%%%

\subsubsection{The Generalized Direct Fitting Procedure}

The final form of the computation of $\chi^2$ is now expressed as
\begin{equation}
\chi^2=|\{G-[P_M (B\circ T) + P_N + \sum_{j=0}^{K}a_j H_j]\}\times
W|^2.
\label{chi2}
\end{equation}
Here, we have written the galaxy spectrum as $G$ and the broadened,
velocity shifted, template spectrum is written as $B\circ T$. Once
again, we explicitly include $W$, the vector of pixel weights. The
coefficients of the $M$-order Legendre polynomial, $P_M$, are solved
for simultaneously along with $P_N$ and $a_j$. The additive continuum
functions can be a polynomial ($P_N$) of order $N$, or a collection of
sines and cosines ($H_j$) up to order $K$.

A simple gradient-search algorithm is employed to search for the values
of $\sigma$ and $V$ which minimize $\chi^2$, which includes weighting
by the inverse of the expected noise. One begins with an approximate
velocity dispersion, $\sigma$, and radial velocity, $V$ and uses these
to broaden and shift the stellar spectrum, $T$. This convolution and
velocity shift can be performed in the Fourier domain. For a given
$\sigma$ and $V$, it is trivial to compute $P_M$, $P_N$, and $a_j$,
taking into account the pixel weighting. In this algorithm, the galaxy
spectrum remains unfiltered in any way to preserve its noise
characteristics and is used directly in the computation of $\chi^2$.
When the $\chi^2$ minimum is found, the errors in dispersion and
velocity are determined from the local topology of $\chi^2(\sigma,V)$.

We use $M=5$ to remove the multiplicative shape differences between
the template and galaxy spectra. Using $M=4$ or $M=6$ has almost no
effect on the resulting velocity dispersions, less than a percent on
average. In tests with $M=0$, we found that the resulting velocity
dispersions were systematically higher, but by $\simlt 5\%$. When the
multiplicative polynomial is included in the fit, the reduced $\chi^2$
values were 25\% lower than when additive functions were used alone.
This dramatic effect is due to the fact that our template spectra have
a very strong slope from blue to red as a result of the grating
response function. Matching the overall shapes of the templates with
those of the galaxy spectra is clearly an important step in the
procedures. For work on distant galaxies, the difference between the
instrumental response function of the galaxy spectra and the templates
should not be removed using additive functions alone. By incorporating
the renormalization into velocity dispersion fitting procedure, we can
be confident that all instrumental effects have been accounted for,
and their effects on the results have been minimized in a
least-squares sense. Given the above test, we are confident that
differences in the instrumental setup lead to systematic errors which
are $\ll 5\%$, and are probably of order $\about 1\%$.

The parameter $K$ is equivalent to the $f_{\rm low}$ parameter of the
Fourier Fitting Method when one uses sines and cosines as the basis
functions for the additive continuum components. As discussed earlier,
$f_{\rm low}$ is the half-power point of the filter in the Fourier
domain, and the effective number of frequencies removed from the
$\chi^2$ sum is $K = 1\case{1}{2} f_{\rm low}$. For these CL1358+62
data, we use $K=13$. For consistency with this filtering, the number
of polynomials one must use in the continuum fitting is defined by the
number of zeroes in the sine and cosine filter. Since each sine or
cosine of frequency $f$ has $2 f$ zeroes, $N = 3 f_{\rm low}$. The
results are insensitive to the choice of $N$ at the level of a few
percent, and merely reflect our susceptibility to template (continuum)
mismatch, which is discussed below.

%%%%%%%%%%%%%%%%%%%%%%%%%%%%%%%%%%%%%%%%%%%%%%%%%%%%%%%%%%%%%%%%%%%%%%%%

\subsubsection{Expanded Variants of the Direct Fitting Method}
\label{otherforms}

Additional terms can be added to Equation \ref{chi2} for a variety of
specialized cases. For example, multiple template spectra can be
incorporated straightforwardly:
\begin{equation}
\chi^2=|\{G- [P_M \sum_{k}^{N_t}\gamma_k (B_k\circ T_k) + P_N +
\sum_{j=0}^{K}a_j H_j]\}\times W|^2.
\end{equation}
In such a case, the fitting is complicated by the requirement of
positive definite solutions for all $\gamma_k$, the relative
contribution of each template, $T_k$. Each template may have its own
broadening function $B_k$, such that each component has its own radial
velocity and velocity dispersion..

Furthermore, for data in which subtraction of the sky background is a
difficult procedure, a high $S/N$ sky spectrum, $S$, may be
incorporated into a fit directly to the raw data, $D$. Thus, the raw
data are approximated by a a sum of the sky with the broadened
template spectrum:
\begin{equation}
\chi^2=|\{D - [P_L S + P_M (B\circ T) + P_N + \sum_{j=0}^{K}a_j
H_j]\}\times W|^2.
\end{equation}
In this case, instead of a sky-subtraction being performed on an
unconstrained column by column basis, as is nominally done, each pixel in
the spectrum contributes to the fit. The sky spectrum is scaled by a
low-order polynomial $P_L$ of order $L$. In the simplest case, $P_L$ can be
a constant, appropriate when large-scale flat-field variations have been
accurately removed. However, when large-scale flat-fielding may be an
issue, the order $L$ should be increased. One can also fit for small
shifts in the the sky vector if small rectification errors remain in
the data. Such an application may prove useful when working with
fiber-fed spectrographs, where determination of an accurate local sky
spectrum is problematic, or when analyzing two-dimensional spectra of
the low surface brightness halos of cD galaxies (\cite{kelsoncd}).

%%%%%%%%%%%%%%%%%%%%%%%%%%%%%%%%%%%%%%%%%%%%%%%%%%%%%%%%%%%%

\section{Tests Between Fitting Methods}
\label{compare}

Our new direct fitting method has been tested against another direct
fitting algorithm, a direct analog of the Franx \etal\ (1989) Fourier
Fitting Method which was used by van Dokkum \etal\ (1998b) to measure
velocity dispersions for galaxies in the cluster MS1054--03. They used
a program analogous to the Fourier Fitting Method, though its
treatment of the continuum filtering is not given by the smooth cosine
bell, but has a rigid cutoff instead. This treatment is similar to the
Rix \& White (1992) prescription for treatment of the continuum for
the case where the continuum functions are sines and cosines.

Since our template spectra had different shapes than our galaxy
spectra, due to the overall efficiency of the spectrograph decreasing
to the blue, we were required to re-normalize the spectrum of the
stellar template to the overall shapes of the galaxy spectra. The new
methodology which we described in \S \ref{disp} already includes a
multiplicative polynomial in the fitting process. The Rix \& White
methodology does not, nor does the Franx real-fitting variant of the
Fourier Fitting method since these programs only use additive
continuum functions in the fitting procedures.

\begin{figure}
\mkfigbox{3in}{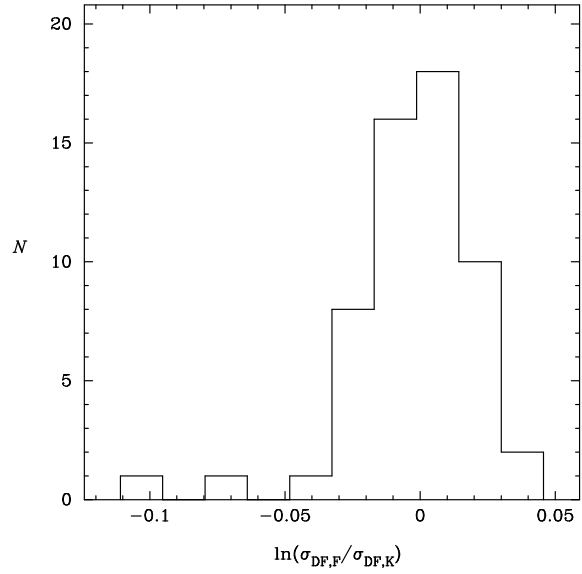}
\caption[fig5.ps]{
A histogram of the differences in $\log \sigma$ between the adopted
velocity dispersions, $\sigma_{\rm DF,K}$, derived using the fitting
method in \S \ref{direct}, and velocity dispersions derived using a
real-fitting variant of the Fourier Fitting method, $\sigma_{\rm
DF,F}$. The data used are listed in in Table \ref{sigfinal}. The
median offset is $ < 1\%$.
\label{realcompare}}
\end{figure}

We have compared our new direct fitting algorithm to the real variant
of the Fourier Fitting method. The aperture-corrected values (\S
\ref{final}) are listed in Table \ref{sigfinal}, in which the results
from the method of \S \ref{direct} are referred to using $\sigma_{\rm
DF,K}$ and values derived using a real-fitting variant of the Fourier
Fitting method are referred to by $\sigma_{\rm DF,F}$. After
renormalizing the templates to the shapes of the galaxy spectra, and
deriving new values of $\sigma$ with Franx's real-fitting variant of
the Fourier Fitting method, we found a mean (median) offset of
$-0.6\%$ ($-0.4\%$) compared to the results of the new direct fitting
algorithm described above, with a standard deviation of 2\%. This
comparison is illustrated in Figure \ref{realcompare}, in which we
show a histogram of the difference between $\log \sigma$ using the two
codes. Another very important consistency check is the comparison of
the reported formal errors, $\delta_\sigma$. The two programs report
the same formal errors in the mean to $< 1\%$ of the error, with a
standard deviation of 5\% of the error.

In order to compare results from our new direct fitting method with
that of the original Fourier Fitting Method, we turned off all pixel
masking in the direct-fitting procedure. Thus, any strong residuals
due to sky-subtraction needed to be interpolated over to ensure that
they did not strongly affect the velocity dispersion fitting,
corresponding to the usual procedures for Fourier fitting.

\begin{figure}
\mkfigbox{3in}{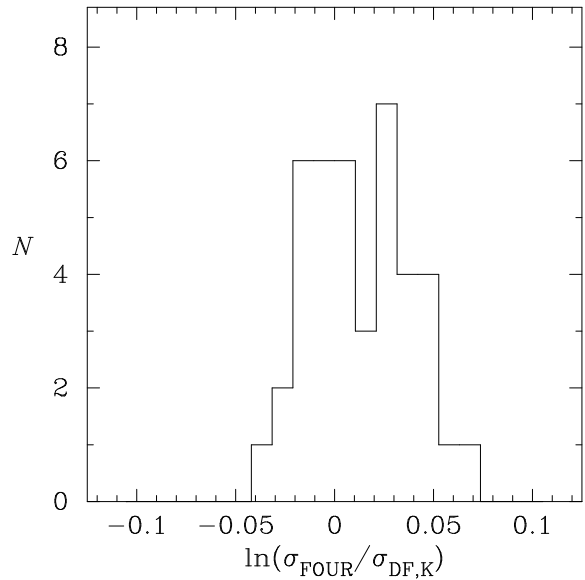}
\caption[fig6.ps]{
A histogram of the differences in $\log \sigma$ between dispersions
derived using the fitting method in \S \ref{direct}, and velocity
dispersions derived using the Fourier Fitting method of Franx \etal\
(1989). The comparison shown was made using the adopted template
HD72324, and uniform pixel weighting. As in Fig. \ref{realcompare},
the median offset is $< 1\%$. We conclude that velocity dispersion
derived using the techniques in \S \ref{direct} can be compare
directly to measurements made using the Fourier Fitting method on
nearby galaxies.
\label{fourcompare}}
\end{figure}

For templates with spectral types later than solar, agreement is quite
good, with very small systematic differences between velocity
dispersions measured with the two algorithms. For most of the stars,
the agreement is at the level of $\simlt 1\%$, with a typical scatter
of 3-4\%. Such offsets can arise from differences in the way in which
the stellar continuum is filtered and matched to the galaxy spectra.
The continuum filtering is an important part of minimizing the
mismatch between the broadened stellar templates and the galaxy
spectra, and any differences in this process will be most important
for stars which already provide poor matches to the galaxy spectra. In
Figure \ref{fourcompare} we show the comparison using the adopted
template HD72324, for those galaxies which required no masking of
emission or excess Balmer absorption.

After these tests, we conclude that the velocity dispersion
measurements made using the direct fitting method are comparable to
those made with both the classical Fourier Fitting Method and its
real-fitting variant, as used by van Dokkum \etal\ (1998b).

%%%%%%%%%%%%%%%%%%%%%%%%%%%%%%%%%%%%%%%%%%%%%%%%%%%%%%%%%%%%%%%%%%%%%%%%

\section{Minimizing Template Mismatch}
\label{select}

\subsection{Selecting the Best Stellar Template}

The fitting procedure(s) outlined in the previous sections provide two
diagnostics for determining which template star best matches the
galaxy data. First, the direct fitting procedure produces a fitted
template, which allows one to visually inspect the residuals and the
quality of the fit. In general, visual inspection allows one to sort
out the best two or three templates. The $\chi^2$ determinations allow
us to determine the best-fit template quantitatively, rather than
visually.

Many of the galaxies show enhanced Balmer absorption, two show
emission, and sharp sky-line residuals exist in nearly all of the
galaxy spectra ({\it e.g.\/} 5577 \AA). Therefore, for every galaxy,
we chose to consistently mask the regions around Na and [O I], the
strongest sky-line residuals in the spectral regions fitted. Some
galaxies also required the masking of strong Balmer lines. The E+As
({\it e.g.\/} ID\# 209, 328, and 343) have strong, broad H$\beta$ and
H$\gamma$ absorption which are not well-matched by any of our current
set of template spectra. The emission line Sb galaxy ID\# 234 and the
cD galaxy ID\# 375 required the masking of the H$\beta$ and H$\gamma$
emission features, as well as [O III].

The star with the lowest median $\chi^2$, overall, is HD72324 (G9III).
We therefore adopt the velocity dispersions found using the template
HD72324, regardless of any individual galaxy's lowest $\chi^2$. The
star which gave the second lowest $\chi^2$ was HD102494 (G9IV).
Dispersions derived using this star were only offset systematically
from values derived using HD72324 by 1\%.

Below each galaxy spectrum in Figure \ref{spectra} we plot the
residuals from the fit of the template HD72324. Regions which have
been masked because of sky line residuals are also masked from these
plots. Regions of the spectra which were excluded from the fitting due
to emission or abnormally strong Balmer absorption are shown in the
figure, and the reader can easily see that the galaxies possess a wide
range of stellar population and star-formation properties. The Balmer
and metal line characteristics of the sample will be analyzed in
detail in a later paper (\cite{kelson99c}).

There still remains the possibility of residual template mismatch {\it
within\/} the sample, artificially inflating the scatter in the
fundamental plane. To test the magnitude of this effect, we compared
dispersions derived using HD72324, to the set of dispersions found
using the lowest $\chi^2$ on an individual basis. For those galaxies
which do not have HD72324 as best-fit template, the mean offset is
$1\%$, with a standard deviation of $2.8\%$. Since these galaxies make
up less than half the sample, the effect is actually much smaller over
the entire sample.

%%%%%%%%%%%%%%%%%%%%%%%%%%%%%%%%%%%%%%%%%%%%%%%%%%%%%%%%%%%%%%%%%%%%%%%%

\subsection{Template Mismatch}

Our method of determining velocity dispersions explicitly assumes that
a single stellar spectrum can provide a good match to any given galaxy
spectra. In general, one should not expect any single star to give a
perfect fit, since galaxy spectra are comprised of the spectra of a
range of stellar spectral types. Some of the mismatch between the
stellar template and the galaxy spectrum is compensated by the
continuum matching described earlier. Remaining mismatch over spatial
frequencies smaller than the filter window, both in the continuum, and
in the detailed depth and shapes of the absorption lines, can be
fitted using composite templates. While the fit can be improved
significantly (\cite{rix92}), this effect is, at most, a few percent
(\cite{jesus}).

Discrepant absorption line depths and small-scale power in the
continuum will depend on several physical parameters, the overall
chemical and spectral makeup of a galaxy itself, the metal abundances
in the star from which the template spectrum was obtained, the
spectral type of the star, etc. Therefore, one might expect that the
velocity dispersions one finds will vary, systematically, with the
spectral type and metallicity of the adopted template (see, {\it
e.g.\/}, \cite{jesus}).

%%%%%%%%%%%%%%%%%%%%%%%%%%%%%%%%%%%%%%%%%%%%%%%%%%%%%%%%%%%%%%%%%%%%%%%%

\subsubsection{Spectral Type}

\begin{figure}
\mkfigbox{4in}{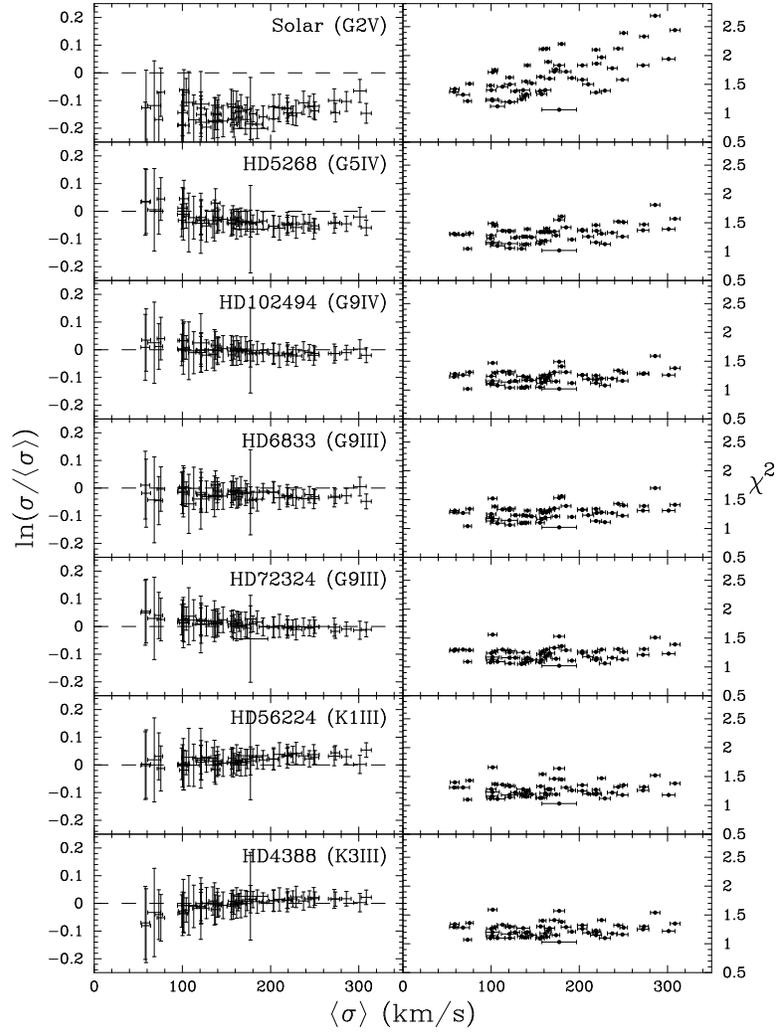}
\figcaption[fig7.ps]{
(a) A comparison of velocity dispersions derived using different template
stars. These measurements were made using the direct fitting algorithm
outlined in \S \ref{direct}. The dispersions derived from each
template are compared here to the average of the dispersions derived
from the five G9-K3 templates. (b) The reduced $\chi^2$ of the fit for
each template as a function of velocity dispersion. For galaxy \#360,
$\chi^2$ does not vary greatly with template because of poor $S/N$.
\label{mismatch}}
\end{figure}

Figure \ref{mismatch} shows the sensitivity of the measured velocity
dispersions for our galaxies to the template's spectral type in the
direct-fitting method. The figure shows how the variation of $\sigma$
with template spectral type is systematic. The separate panels show
the logarithmic difference between each of the templates and the
average of the five G9-K3 stars, chosen as a reference. The most
deviant template is that of the solar spectrum (G2V). Using the solar
spectrum as a template gives dispersions which are quite deviant, 15\%
lower on average, systematically. The next worse template is the G5IV
star HD5268, which gives values 4\% lower than those obtained with the
best fit template HD72324.

A variation of mismatch with dispersion can be a worry for fundamental
plane work, since, in principle, the slope with respect to $\log
\sigma$ can be biased. In the extreme case of the solar spectrum, the
offset in $\sigma$ from the values obtained from late-type stars
differs between a 100\kms\ galaxy and a 300 \kms\ galaxy by about
$\Delta \log(\sigma) \approx 0.1$; the slope of the FP is then biased
by $\delta \alpha \approx 0.2$. However, the stars with the lowest
$\chi^2$ values yield values for $\sigma$ which systematically differ
by at most a percent or two over such a long baseline of $\sigma$.
Thus, the impact on the fundamental plane slope is only a few percent.

%Such template mismatch would not likely introduce much scatter into
%the fundamental plane because the scatter in $\sigma$ in the
%template-to-template comparisons remains consistent with the
%observational uncertainties (aside from the template of the solar
%spectrum).

%%%%%%%%%%%%%%%%%%%%%%%%%%%%%%%%%%%%%%%%%%%%%%%%%%%%%%%%%%%%%%%%%%%%%%%%

\subsubsection{Metal Abundance}

Template mismatch is also a function of the absorption line strengths in
the stellar spectra (at a given spectral type). Two of our stars, the
two G9III stars, have metallicities differing by a factor of 10
(\cite{luck,pila}). To zeroth-order, this difference is compensated for
by $\gamma$, the line-strength parameter in the velocity dispersion
fits. This compensations is clearly seen in the case of these two stars,
as the average ratio of $\gamma_{\rm HD72324}/\gamma_{\rm HD6833}\approx
0.8$. These two stars with identical spectral type, and a factor of 10
difference in metal abundances, show a 3\% systematic offset in velocity
dispersion. They also show a 3\% systematic trend in $\sigma$ over a
baseline of 0.5 dex in $\sigma$, but only a 2\% trend when one excludes
those galaxies with $\sigma < 100$ \kms. By choosing the best-fit
template, one's systematic template mismatch errors should therefore be
smaller still. By simply choosing a late-type star, one is already
reducing the mismatch errors to a few percent.

Procedures have been developed which combine multiple template spectra
to create an optimal match to the galaxies ({\it e.g.\/},
\cite{rix92}). Such algorithms adjust the mix ({\it i.e.\/}, $\gamma$)
of stars to create a non-negative distribution of stellar components.
The direct-fitting method can be expanded fit for an optimal
combination of template spectra, as shown in \S \ref{otherforms} but
such an algorithm is not necessary for our purposes. Our uncertainties
due to template mismatch are already small, at a level comparable to
the formal errors from the velocity dispersion fits.

%%%%%%%%%%%%%%%%%%%%%%%%%%%%%%%%%%%%%%%%%%%%%%%%%%%%%%%%%%%%%%%%%%%%%%%%

\subsection{Wavelength-Dependent Mismatch}

There is an additional important test which serves two purposes. By
comparing velocity dispersions derived from different sections of the
galaxy spectra one can determine if (1) one has indeed used a template
which best matches the galaxy spectra globally; and if (2) the
measured velocity dispersions depend on the spectral region used in
the fit. The latter is an important question since our fitting is
targeted at a region of the spectrum containing the G band; we wish to
know if our measurements can be compared to velocity dispersions
measured in other regions of galaxy spectra ({\it e.g.\/}, Mg $b$).

If one is using a template which provides a perfect match to a given
galaxy spectrum, then one should measure identical values for $\sigma$
using any region of the spectrum. The practical approach to this is to
test if the dispersion measured from the blue side of the spectrum
equals that measured from the red side. Therefore, we split our galaxy
spectra in two, and measured $\sigma$ from the two halves, for the 46
galaxy spectra with $S/N \ge 10$ per pixel in {\it both\/} halves.

\begin{figure}
\mkfigbox{4in}{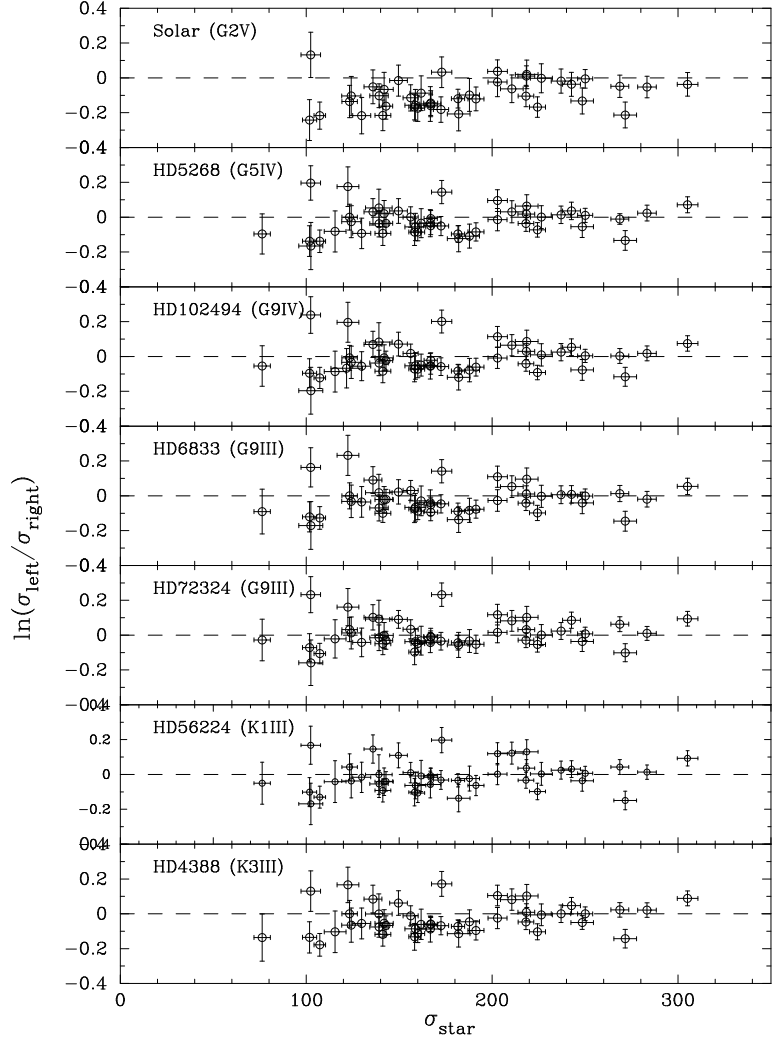}
\figcaption[fig8.ps]{
A comparison of $\sigma$ measured from the left and right halves of each
spectrum. The difference in dispersions measured from the left and
right halves of the galaxy spectra are shown as a function of $\sigma$
for a given template. The direct fitting method of \S \ref{direct} was
used, with pixel-weighting and masking wherever necessary.
\label{sides}}
\end{figure}

Figure \ref{sides} shows the logarithmic difference in $\sigma$ from
the two sides for each template. The best-fit template star, HD72324,
yields a mean difference in $\sigma$ between the two halves of less
than a percent. In fact, most of the template stars give reasonable
results. The rest of the stars, aside from the solar spectrum, all
show absolute differences on the order or less than a few percent. The
solar spectrum shows a median difference between the two halves of
$-10\%$.

\bigskip

As a result of these multiple tests, we are confident that our errors
due to template mismatch are likely to be small, at the level of a few
percent, comparable to the formal errors in the fit.

%%%%%%%%%%%%%%%%%%%%%%%%%%%%%%%%%%%%%%%%%%%%%%%%%%%%%%%%%%%%%%%%%%%%%%%%
%%%%%%%%%%%%%%%%%%%%%%%%%%%%%%%%%%%%%%%%%%%%%%%%%%%%%%%%%%%%%%%%%%%%%%%%
%%%%%%%%%%%%%%%%%%%%%%%%%%%%%%%%%%%%%%%%%%%%%%%%%%%%%%%%%%%%%%%%%%%%%%%%

\section{The Velocity Dispersions}
\label{final}

Early-type galaxies have radial gradients in velocity dispersion and
radial velocity. The implication is that the $\sigma$ one measures
depends upon the metric aperture used to extract the original spectra.
The aperture is defined by an angular extent upon the sky, so the
metric aperture which defines the velocity dispersion typically
increases with distance to a given galaxy.

The dependence of $\sigma$ upon aperture is not straightforward to
predict because the measurement is a luminosity-weighted mean velocity
dispersion within the predefined aperture. Furthermore, early-type
galaxies can be partially supported by rotation, and velocity
dispersion measurements made using large metric apertures will reflect
the total second moment of the integrated line-of-sight velocity
distribution: $\mu_2\equiv \sqrt{\sigma^2 + V^2}$. The integration of
this quantity is weighted by the surface brightness distribution of
the galaxy within the aperture of the spectrograph. Thus, measurements
of $\sigma $ depend on each galaxy's intrinsic distribution of
orbits, as well as its light distribution.

%%%%%%%%%%%%%%%%%%%%%%%%%%%%%%%%%%%%%%%%%%%%%%%%%%%%%%%%%%%%

\subsection{Spatially Resolved Kinematics}
\label{space}

Our velocity dispersion measurements will be compared directly with
similar measurements made with nearby galaxies, {\it e.g.\/}
J\o{}rgensen \etal\ (1996). Therefore, one source of concern is that
the galaxies in the CL1358+62 sample may have different internal
structure than nearby galaxies. We begin exploring the magnitude of
this effect by comparing the kinematic profiles of our galaxies with
those of nearby galaxies.

With the excellent image quality and high $S/N$ of the data, we were
able to extract rotation curves and velocity dispersion profiles.
Using the direct-fitting method of \S \ref{direct} we measured $V$ and
$\sigma$ along each slitlet. When necessary, rows were binned to
achieve sufficient $S/N$. The $\chi^2$ sum used the same masking and
noise-weighting as in the determination of the $1\square ''$ aperture
dispersions. Two of the galaxies have $S/N$ ratios too poor to provide
spatially resolved kinematics (\# 360 and \# 493).

The profiles can be seen in Figure \ref{kinprof}, where absorption
line rotation curves and velocity dispersion profiles are plotted
adjacently for each galaxy. These profiles have not been corrected for
seeing, nor for non-major axis positioning of the slitlets. For most
high-redshift galaxies, the rise in the rotation curve has been
heavily modified by seeing (see, for example, the modeling of
\cite{vogt,vogt97}).

%%%%%%%%%%%%%%%%%%%%%%%%%%%%%%%%%%%%%%%%%%%%%%%%%%%%%%%%%%%%%%%%%%%%%%%%

\subsection{Comparison with Nearby Galaxies}
\label{dispcor}

\begin{figure}
\mkfigbox{5in}{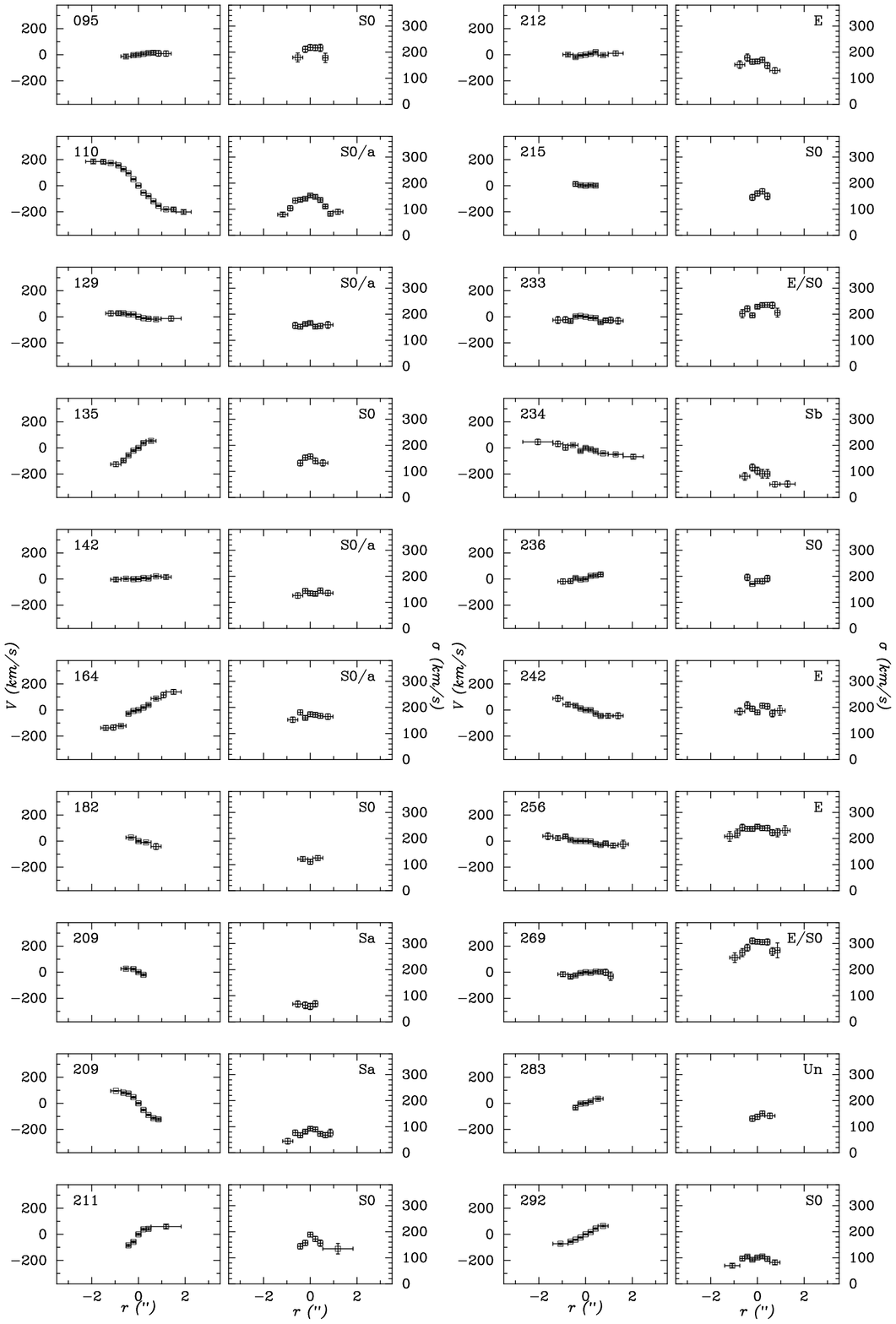}
\figcaption[fig9a.ps,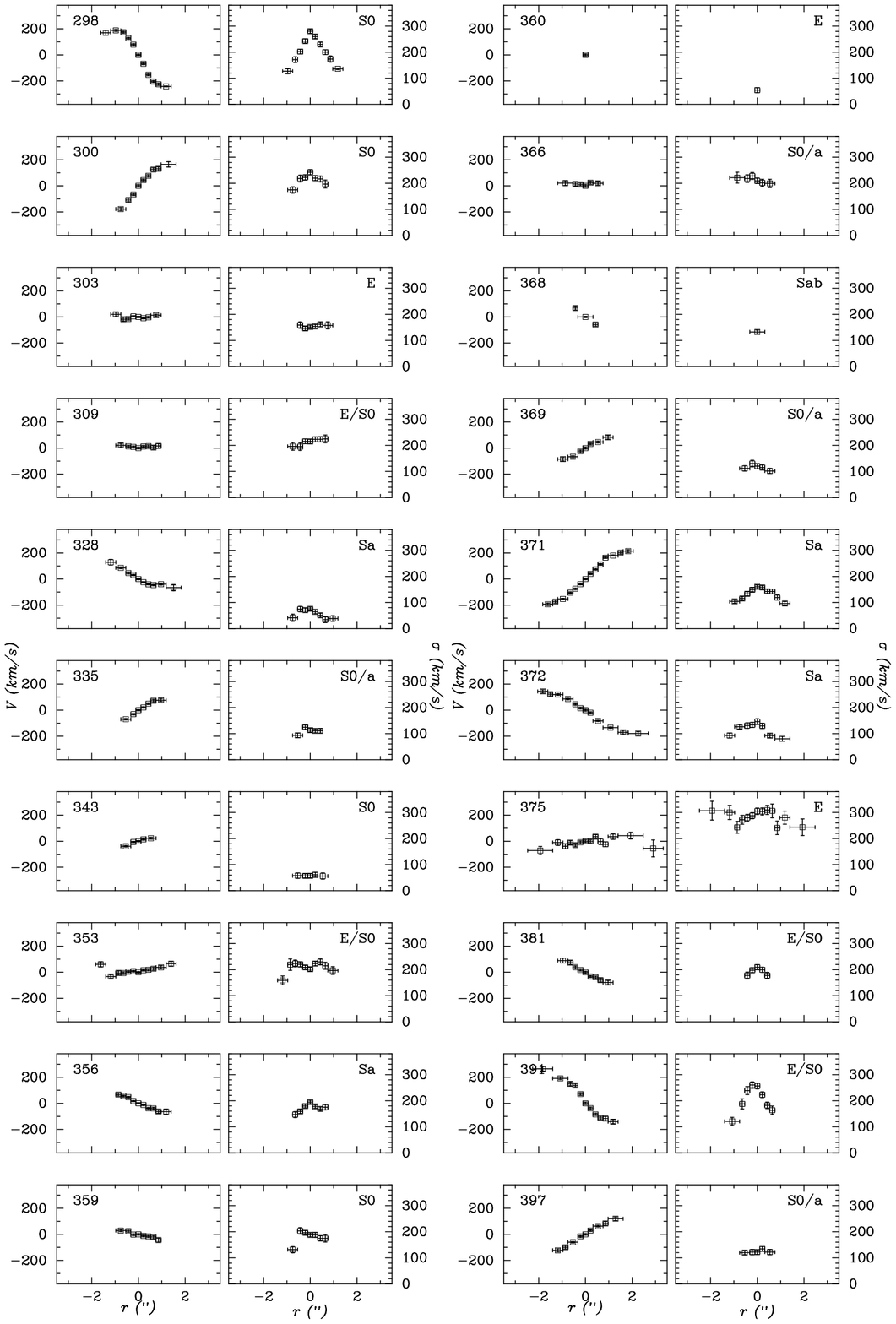,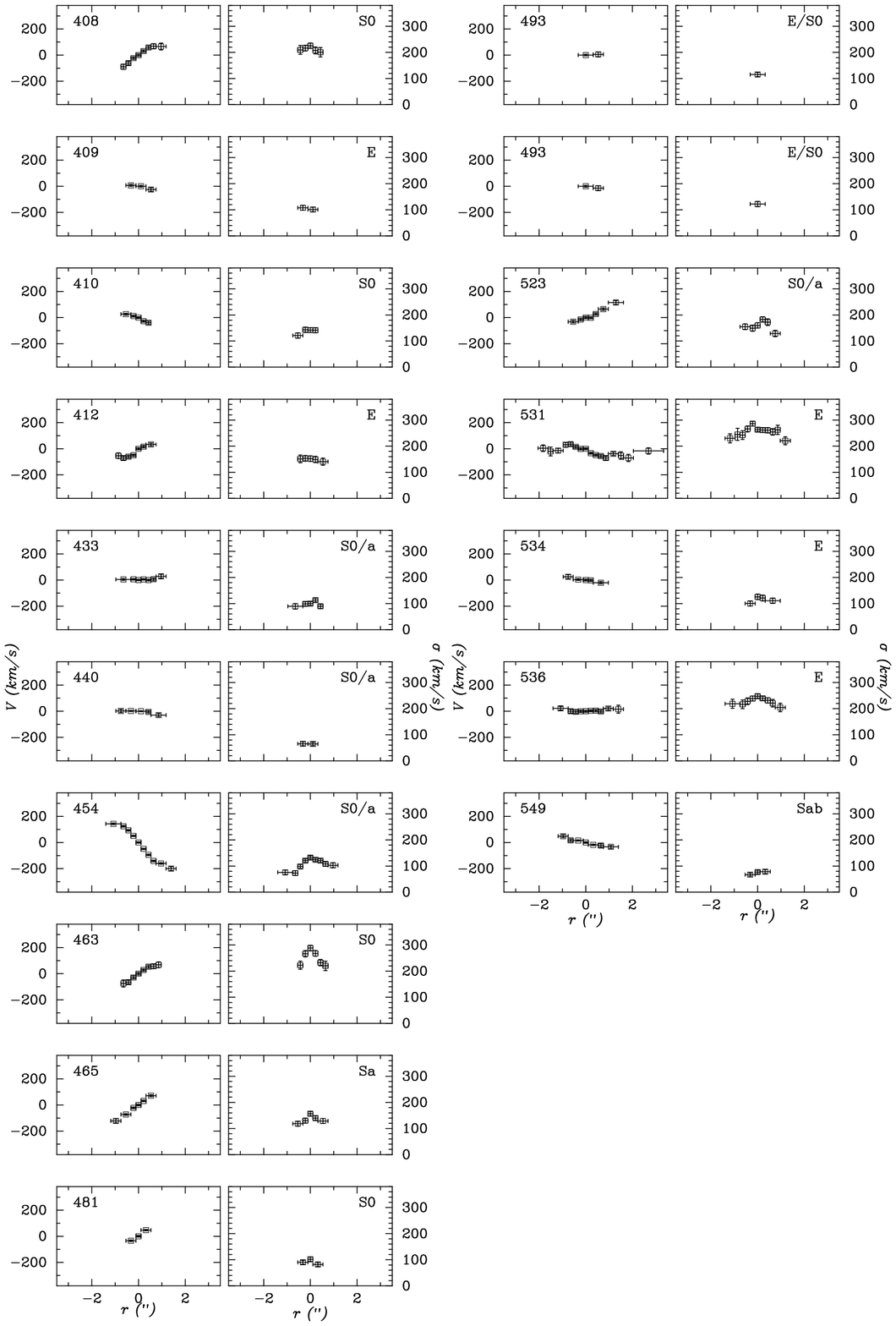]{
Spatial profiles of $V$, and $\sigma$. These have not been corrected
for seeing effects, or for inclination.
\label{kinprof}}
\end{figure}

\begin{figure}
\mkfigbox{3.5in}{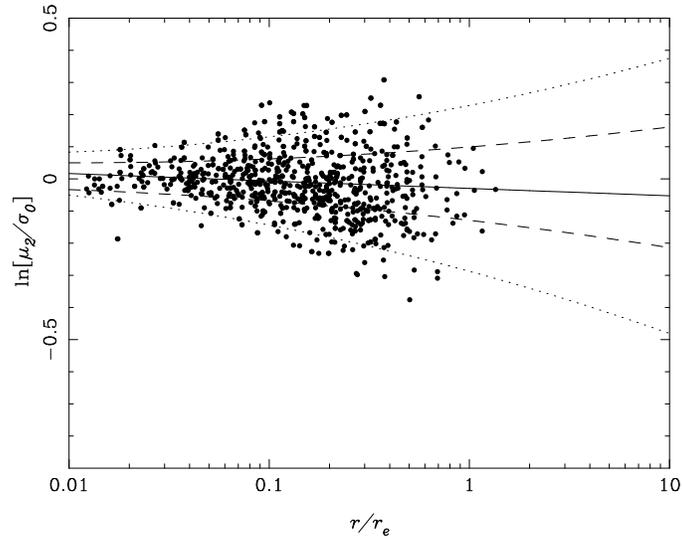}
\figcaption[fig10.ps]{
The logarithmic gradient of the second moment of the kinematics,
$\mu_2=\sqrt{\sigma^2+V^2}$, of nearby galaxies using long-slit
observations from the literature (\cite{sp97a},b,c). A first-order
least-squares fit to these data is shown as the solid line. A second
order least-squares fit to the \rms\ scatter is used to show the $\pm
1\sigma$ and $\pm 2\sigma$ variation with radius along galaxy
major-axis (the dashed lines). The profiles of the CL1358+62 galaxies,
shown in Figure \ref{apcor}, are very similar to those shown here of
nearby galaxies.
\label{localprof}}
\end{figure}

In order to test whether aperture corrections derived from nearby
galaxies can be applied to the high redshift sample, we now compare
the kinematic profiles in Figure \ref{kinprof} to those of nearby
galaxies. For this comparison, we use 83 early-type galaxies in the
kinematic database of Simien \& Prugniel (1997a,b,c) and construct
$\mu_2=\sqrt{\sigma^2+V^2}$ as a function of radius using their
long-slit rotation curves and velocity dispersion profiles. In Figure
\ref{localprof}, we show the logarithmic gradients of $\mu_2$,
normalized by the mean value within $r_e/10$ (we plot only those data
with uncertainties smaller than 10\%). A least-squares fit to the data
in Fig. \ref{localprof} yields a logarithmic gradient of
$d\ln(\mu_2)/d\log r=-0.023\pm 0.006$ per dex, shown by the solid
line. The effective $\pm 1$ and $\pm 2\sigma$ scatter about the fitted
gradient is shown by the dashed and dotted lines, respectively.

\begin{figure}
\mkfigbox{5in}{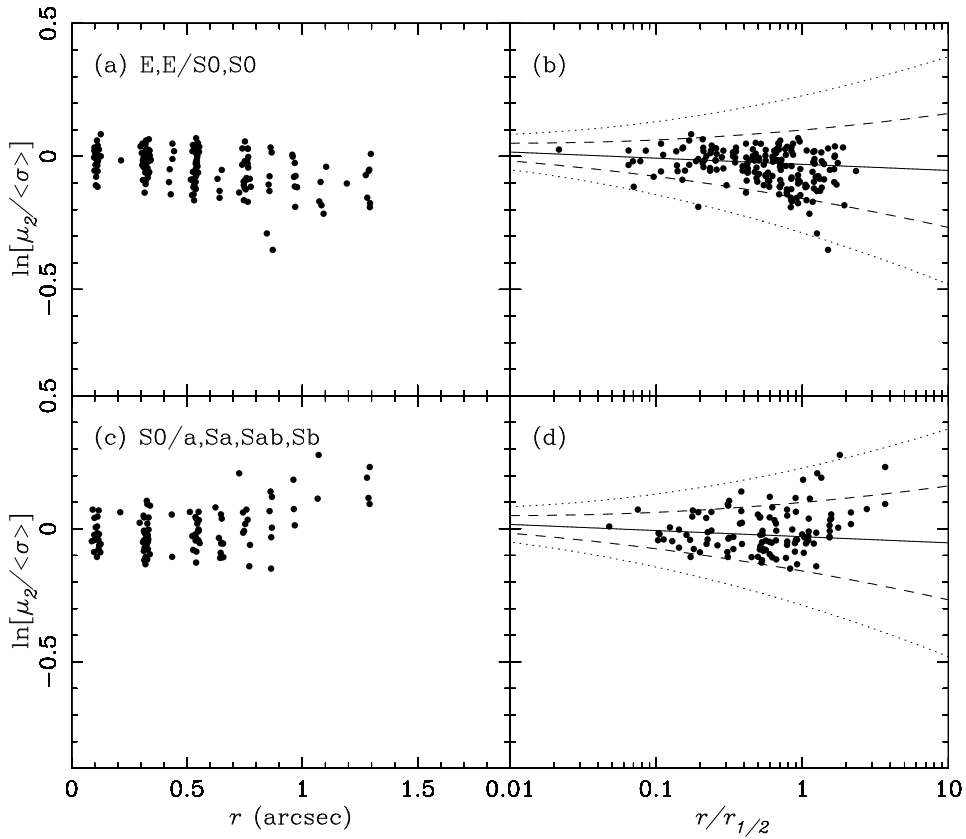}
\figcaption[fig11.ps]{
The logarithmic gradients of the second moment of the kinematics,
$\mu_2=\sqrt{\sigma^2+V^2}$ as functions of radius in arcsec (a,c),
and radius with respect to the half-light radii of the galaxy (b,d;
from bulge-plus-disk profile fitting to HST surface photometry; see
\cite{kelson99a}). These have not been corrected for seeing effects.
Artificial scatter has been added to the radii in (a,c) to aid in
displaying the data. The E/S0 galaxies are shown in the top panels,
and the spirals in the bottom. Within a half-light radius, the
kinematic profiles appear to be independent of morphology. The
least-squares fit to the local data of Fig. \ref{localprof} are shown
here in (b,d) and indicate that the kinematic gradients of the
CL1358+62 galaxies are similar to nearby galaxies.
\label{apcor}}
\end{figure}

In Figure \ref{apcor}(a,c), we plot the profiles of
$\sqrt{\sigma^2+V^2}$, constructed from the data in Figure
\ref{kinprof}, for all of the CL1358+62 galaxies. As in Figure
\ref{localprof}, we plot only those data with uncertainties smaller
than 10\%. These gradients have not been corrected for seeing. In
(b,d) we show the gradients, normalizing the position along the slit
by the galaxy half-light radii, taken from Kelson \etal\ (1999a). In
these figures we plot the first-order fit from Fig. \ref{localprof} as
the solid line. We show the $\pm1\sigma$ (dashed) and $\pm2\sigma$
(dotted) contours of the nearby galaxies as well. A least-squares fit
to all of the CL1358+62 galaxies yields a logarithmic gradient of
$d\ln(\mu_2)/d\log r=-0.017\pm 0.026$ per dex, consistent with the
sample of nearby galaxies.

From the figures, we draw several important conclusions. First, the
$\mu_2$ profiles of E/S0s and early-type spirals have small gradients,
both locally, and at $z=0.33$. Second, within $r<r_{1/2}$, the
gradients are independent of morphology. Least-squares fits to the
gradients in the spirals are not significantly different from fits to
the gradients in the E/S0s. The scatter in the $\mu_2$ profiles at
$z=0.33$ is remarkably consistent with that in the nearby galaxies. We
therefore conclude that aperture corrections derived from nearby
galaxies can safely be used on the full sample of CL1358+62 galaxies.

%%%%%%%%%%%%%%%%%%%%%%%%%%%%%%%%%%%%%%%%%%%%%%%%%%%%%%%%%%%%

\subsection{Aperture Corrections}

J\o{}rgensen \etal\ (1995) used existing spectroscopic and photometric
data of nearby galaxies in the literature to construct models which
were ``observed'' through a series of concentric apertures of
increasing size. Major-axis long-slit slit spectroscopy, by itself, is
not suitable for deriving such corrections. They found that the
following power law can be used to correct the observed dispersions,
$\sigma_z$, to a physical aperture:
\begin{equation}
\log \sigma_c = \log \sigma_z + 0.04 (\log d_z - \log d_c),
\label{eq:apcor}
\end{equation}
where $\sigma_c$ is the corrected velocity dispersions within an
aperture $d_c$, and $d_z$ is the aperture of the observation. For this
paper, we use $d_c$ as the nominal aperture of $3 \Sec 4$ for galaxies
at the distance of Coma, and $d_z$ is the effective aperture for
CL1358+62 of $1\Sec 23$ (see \S \ref{data}).

We used Eq. \ref{eq:apcor} to correct the observed velocity
dispersions to an aperture consistent with the comparison sample of
J\o{}rgensen \etal\ (1996). Using $q_0=0.5$, we obtain values from the
standard aperture at Coma of $d_c=1.68$ kpc, and an aperture size of
$d_z=5.50$ kpc at $z=0.33$. Therefore, our measured velocity
dispersions underestimate the central values by $4.7\%$. Most of the
slitlets were not tilted and thus do not suffer additional broadening.
The most extreme case of a $45^\circ$ tilt (see \S \ref{data}) changes
the aperture correction to $5.4\%$. Thus, we adopt a single aperture
correction, regardless of slit tilt. The final velocity dispersions
have therefore been multiplied by a factor of $1.047$.

\begin{figure}
\mkfigbox{2.5in}{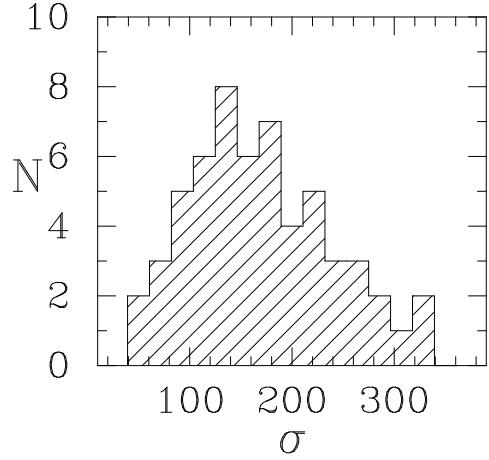}
\figcaption[fig12.ps]{
Histogram of measured velocity dispersions.
\label{sighist}}
\end{figure}

The velocity dispersions, corrected to an aperture equivalent to the
nominal aperture at the distance of Coma, are listed in Table
\ref{sigfinal}. A histogram of central velocity dispersions is given
in Figure \ref{sighist}.

%%%%%%%%%%%%%%%%%%%%%%%%%%%%%%%%%%%%%%%%%%%%%%%%%%%%%%%%%%%%%%%%%%%%%%%%

\section{Sources of Uncertainty}
\label{errors}

When measuring velocity dispersions potential sources of error can
exist at several stages in the procedures. Uncertainties are
introduced by the templates themselves, as well as by the galaxy
spectra. Poor $S/N$ clearly introduces large random errors, but can
also introduce systematic errors (\cite{jfk95}). Furthermore, the
imposition of a single stellar component on a galaxy spectrum, of any
$S/N$, can produce mismatch errors. We attempt to itemize these
sources of error in a quantitative manner, and estimate to what extent
our measurements are affected by them.

%%%%%%%%%%%%%%%%%%%%%%%%%%%%%%%%%%%%%%%%%%%%%%%%%%%%%%%%%%%%%%%%%%%%%%%%

\subsection{The Instrumental Resolution}

At the outset, a potentially large source of error can be improperly
measured resolution. In \S \ref{prep} we found that the uncertainty in
the absolute resolution of the template spectra is uncertain by no
more than \about 5 \kms. This translates to an error at $\sigma=100$
\kms\ of \about 2.5\%. Since the resolution is more important for low
velocity dispersions, there is almost no effect for galaxies with
$\sigma=300$ \kms. Over the baseline of 100 \kms\ to 300 \kms, the
mean systematic effect on the velocity dispersions is of order 1-2\%.
Given that the systematic effect is a function of velocity dispersion,
there is a net uncertainty in the derived fundamental plane slopes
of about $\Delta (\partial \log r_e / \partial \log \sigma) \approx
\pm 0.02$.

%%%%%%%%%%%%%%%%%%%%%%%%%%%%%%%%%%%%%%%%%%%%%%%%%%%%%%%%%%%%%%%%%%%%%%%%

\subsection{Aperture Corrections}

In \S \ref{dispcor}, we showed that the total kinematic profiles of
the Cl1358+62 galaxies is similar to those of nearby galaxies, and
concluded that the J\o{}rgensen \etal\ (1995) form of the aperture
corrections remained valid for our sample. We therefore conclude that
any systematic error in adopting their aperture corrections for the
entire sample is going to be negligible compared to other sources of
error, such as template mismatch. More sophisticated modeling is
required to better estimate this uncertainty.

%%%%%%%%%%%%%%%%%%%%%%%%%%%%%%%%%%%%%%%%%%%%%%%%%%%%%%%%%%%%%%%%%%%%%%%%

\subsection{Signal-to-Noise Considerations}

According to J\o{}rgensen \etal\ (1995), if the $S/N$ ratio of galaxy
spectra are too low, the derived velocity dispersions will be measured
systematically high (see their Figure 3). Five of our spectra have
$S/N \simlt 20$ per \AA\ (two are of the same galaxy, \#493). Only one
has $S/N < 15$ per \AA. It is \#360 with a measured $\sigma = 173 \pm
20$ km/s (a 12\% uncertainty). According to J\o{}rgensen \etal\
(1995), this measurement is in error by only +2\%. For this galaxy,
the low $S/N$ appears to be due to poor subtraction of the background
in a crowded region of the sky. The slitlet containing \#360 is $12''$
long, with \#360 at one end, partially covering the outskirts of the
cD \#375 at the other end, and containing \#368 in between the two.
The other four spectra have measured velocity dispersions between 100
km/s and 140 km/s and, according to the simulations of J\o{}rgensen
\etal\ (1995), may be systematically too large by 1-2\% as well. Given
that these systematic uncertainties are much smaller than the formal
uncertainties, and that only four galaxies, at most, are affected, we
do not include any corrections for this effect.

%%%%%%%%%%%%%%%%%%%%%%%%%%%%%%%%%%%%%%%%%%%%%%%%%%%%%%%%%%%%%%%%%%%%%%%%

\subsection{Template Mismatch vs. Signal-to-Noise}

The formal errors in the velocity dispersions have two primary
components: random errors due to photon statistics and systematic
errors due to, for example, template mismatch. In this section, we
determine to what extent our formal errors are dominated by the random
or systematic errors.

We compare the systematic errors due to template mismatch to the
random errors by comparing three estimators of $S/N$: (1) the $S/N$
ratio per \AA\ expected solely from considerations of the read noise
and noise in the sky and galaxy; (2) $\sigma/\delta_\sigma$, the ratio
of the velocity dispersion with its formal error; and (3) $Q$, the
ratio of the mean flux level in the galaxy and the standard deviation
of the residuals in the velocity dispersion fit. In the case where one
uses a perfect template spectrum and no noise has been added to the
data during the reduction process, $Q$ should be equivalent to the
expected $S/N$. All of these are related (since the expected noise was
used in the calculation of weights in the fitting), and we can use
them to disentangle the contributors to our uncertainties.

\begin{figure}
\mkfigbox{4in}{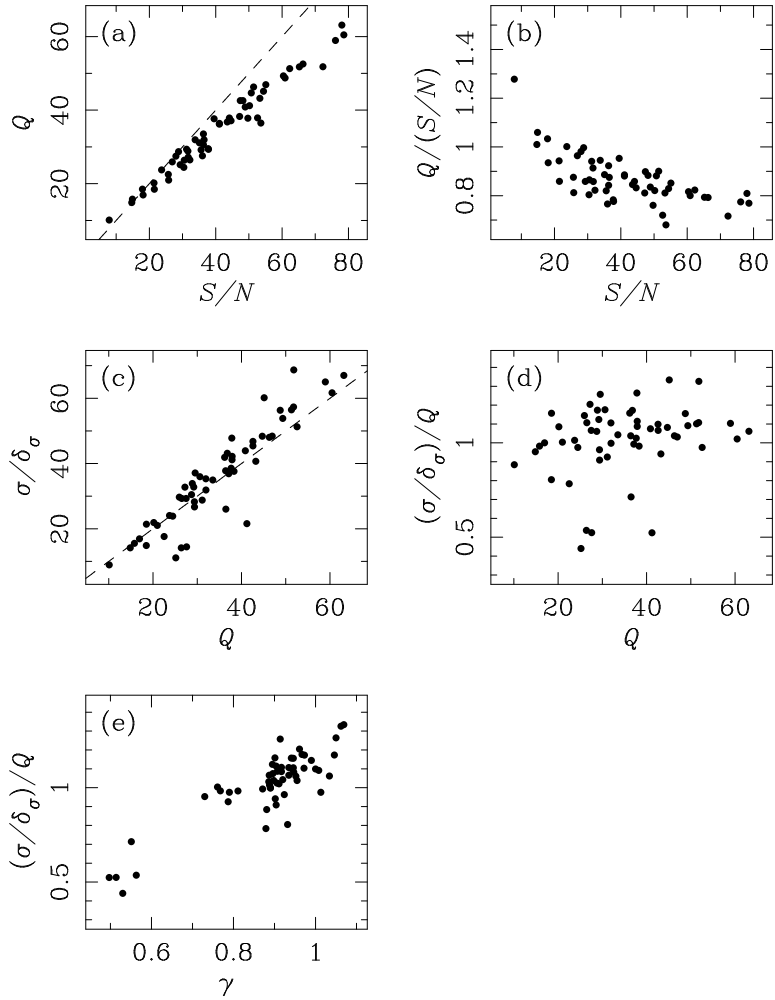}
\figcaption[fig13.ps]{
(a) $Q$, a quality of fit defined as the ratio of galaxy flux to the
standard deviation of the residuals of the velocity dispersion fit, as a
function of the expected $S/N$ per pixel. (b) the ratio of $Q / (S/N)$ as
a function of $S/N$. Note that the higher the $S/N$, the more the
residuals from the velocity dispersion fit become dominated by sources
of error other than the expected noise. In (c), we plot the ratio of
velocity dispersion to its formal error, $\sigma /\delta_\sigma$, as a
function of $Q$, and in (d) and (e), we plot the ratio of
$\sigma/\delta_\sigma$ to $Q$ as functions of $Q$ and $\gamma$,
respectively. The formal uncertainties in the velocity dispersions are
reasonably well given by the quality of the velocity dispersion fit,
though $\gamma$, the line-strength, is clearly a third parameter,
indicating that template mismatch is playing a role. The five data
points in the lower left corner of (e) are the E+A and emission line
galaxies, which are poorly fit by the G9III template star, even when
the Balmer lines are excluded from the fit. In these cases, the formal
errors of the velocity dispersions are higher than one would expect
from the quality of the fit, presumably because of mismatch with the
late-type stellar template.
\label{s2n}}
\end{figure}

In Figure \ref{s2n}(a) we plot $Q$ as a function of the expected $S/N$
ratio (per pixel in this case). The data do not follow the line of
unity slope. For those spectra with high $S/N$, the residuals from the
fit are larger than would be expected from photon statistics. This
point is repeated in Figure \ref{s2n}(b) in which we specifically show
the ratio of $Q/(S/N)$ as a function of expected $S/N$. When the $S/N$
approaches $\about 30$, the template mismatch errors become the
dominant source of uncertainty. Therefore, we do not see a constant
level in (b), but a trend with $S/N$ in which the quality of the fit
begins to suffer compared to what one would expect from noise
considerations alone. This fact can also be seen by inspecting the
residuals from the fits in Figure \ref{spectra}, in which there appear
to be consistent features in the galaxy spectra which are not well
matched by the the fitted template.

In figure \ref{s2n}(c) we plot $\sigma/\delta_\sigma$, the formal
signal-to-noise in the dispersion, as a function of $Q$, the quality
of the fit for each spectrum. For most galaxies, there is clearly a
one-to-one correspondence. In (d), the ratio is plotted as a function
of $Q$ so that one can more clearly see the discrepant data points.
Thus our error estimates are consistent with the \rms\ scatter about
the fit, though some galaxies clear deviate from this expectation.

There are a handful of spectra, however, with velocity dispersion
uncertainties which are larger than expected from the \rms\ residuals
of the fit. We show in in Figure \ref{s2n}(e) that the errors are
dominated by atypically large template mismatch. These are the
galaxies with small values of the the line-strength parameter
$\gamma$. These points with very low $\gamma$ values are galaxies
\#209 (both spectra), \#234, \#328, and \#343, the E+A and
emission-line galaxies. For these cases, the formal errors in the
velocity dispersion are higher than expected for the quality of the
fit of a single template star, even though the strong Balmer
absorption lines and emission lines were given zero weight in the fit.

We conclude that for the galaxies with $S/N > 30$, our velocity
dispersion uncertainties are dominated by systematic errors, such as
template mismatch errors (and as noted, these errors are $\simlt
5\%$).

%%%%%%%%%%%%%%%%%%%%%%%%%%%%%%%%%%%%%%%%%%%%%%%%%%%%%%%%%%%%%%%%%%%%%%%%

\subsection{Internal Consistency}
\label{consist}

For the galaxies in common with Kelson \etal\ (1997), the agreement in
$\sigma$ is 1\% in the mean, with a scatter of 4-5\%, consistent with
the formal uncertainties.

We can estimate the internal consistency of the dataset using two
galaxies which were measured in two different masks. galaxy ID\# 493
was observed at two different position angles, $45^\circ$ apart. The
two velocity dispersions, $117\pm 8$ \kms\ and $128\pm 9$ \kms, agree
very well. The other galaxy, ID\# 209, is an E+A whose spectrum
required masking of the strong Balmer absorption. The two observations
for this galaxy do not agree because the one with the smaller $\sigma$
was made with a slightly mis-aligned slitlet. In this discrepant
observation, the galaxy was a secondary target in a slitlet which was
centered on a different galaxy and subsequently tilted to include \#
209. The slitlet was not correctly tilted to encompass the center of
\# 209, and thus the galaxy was not centered in the slitlet. Thus, the
seeing convolved galaxy did not evenly fill the slit, leaving a strong
gradient of its light profile across the slit. Thus, its light profile
under-filled the slit width, leading to an improper estimate of the
resolution of its spectrum. Furthermore, the resulting dispersion is
not a central value, but one offset from the center. Thus, the larger
value of $\sigma=108\pm 5$ \kms\ is the more credible measurement. No
other galaxies in this sample were secondary targets within slitlets.

%%%%%%%%%%%%%%%%%%%%%%%%%%%%%%%%%%%%%%%%%%%%%%%%%%%%%%%%%%%%%%%%%%%%%%%%

\subsection{The Effects of Under-filling Slitlets}

At these high redshifts, it might be expected that small galaxy sizes,
combined with the good seeing of our observations, might lead us to
under-fill the 1\Sec 05 wide slitlets. If a galaxy under-fills a slitlet,
then the broadening, as measured by the sky-line widths, would be
overestimated, because the galaxy light is coming from an effective
aperture narrower than the slitlet. Fortunately, the galaxies in this
sample are generally not small enough to cause problems. As shown in Kelson
\etal\ (1999a) the smallest galaxy, \#493, has a half-light radius of
\about 0\Sec 2. Using the spatial profile along the slit of this galaxy,
within a box $1\Sec 05$ wide, we have convolved a spectrum of the sky, and
compare the widths of the convolved lines with the widths of the sky lines
convolved by a top-hat $1\Sec 05$ wide. One does find that the sky lines
are narrower in the case where the galaxy profile was used in the
convolution, but the effect is approximately 2\% of the measured widths. We
therefore conclude that under-filling of the slitlets is a negligible source
of error.

\medskip
Given the number of sources of systematic error, we estimate that the
total systematic errors in the order of $\simlt 5\%$, generally larger
than the random errors expected from photon statistics alone

%%%%%%%%%%%%%%%%%%%%%%%%%%%%%%%%%%%%%%%%%%%%%%%%%%%%%%%%%%%%%%%%%%%%%%%%

\section{Summary and Conclusions}

We have accurately measured the internal kinematics of a large,
homogeneous sample of galaxies, within the cluster CL1358+62. These
data will be combined with the structural parameters in Kelson \etal\
(1999a) for the analysis of $M/L$ ratios and the fundamental plane of
early-type galaxies (\cite{kelson99b}. These direct measurement of
velocity dispersions, will provide crucial mass estimates for these
galaxies, enabling us to provide constraints on early-type galaxy
evolution and their epoch(s) of formation. The formal errors are quite
small, typically a few percent. For galaxies with $S/N >30$ (roughly
half the sample), the uncertainties are dominated by the systematic
errors, which appear to be at a level of $\simlt 5\%$.

These velocity dispersion measurements can be compared directly to
those made in nearby galaxies. We have measured $\sigma$ using a new
variant of direct (real) fitting. Results from this method have been
compared with those of the Fourier fitting method (\cite{fih89}), and
we find excellent agreement between the two programs. Differences
between the two algorithms lead to systematic uncertainties in
$\sigma$ on the order of \about $\pm 1\%$, with a scatter consistent
with the formal errors.

We have measured absorption-line kinematic profiles for most of the
galaxies in this sample. The CL1358+62 galaxies show a wide range of
rotational support. However, the profiles of the total second moment
of the LOSVDs, $\mu_2\equiv \sqrt{\sigma^2+V^2}$, do not vary greatly
from galaxy to galaxy. Thus, we conclude that a single aperture
correction can be applied to our entire sample. Furthermore, because
the kinematic profiles of the CL1358+62 galaxies are similar to those
of nearby galaxies, we conclude that we have reliably corrected our
large-metric-aperture velocity dispersions to an effective aperture
consistent with the nearby galaxy samples of J\o{}rgensen \etal\
(1995).

For many of the galaxies, the raw kinematic profiles reach 2-3 $r_e$
(see \cite{kelson99a}). However, due to the effects of seeing and the
large (metric) slit width, sophisticated modeling is required to
derive the true kinematic profiles to such large radii.

E+As appear to be largely supported by rotation. This is consistent
with the presence of disks, as shown in HST imaging of intermediate
redshift E+A galaxies in this, and other clusters
(\cite{franx93,wirth,kelson99a}). These galaxies do not appear to be
massive, having velocity dispersions of $\sigma \simlt 100$ \kms. A
more detailed discussion of these galaxies is given in the context of
the fundamental plane and the $M/L$ ratios of cluster galaxies in
Kelson \etal\ (1999b).

This sample, with more than fifty galaxies in a single cluster, shows
that the new, large-aperture telescopes can provide extremely
accurate, high quality spectroscopic data on distant galaxies. Such
data provide estimates of mass scales, such as those presented here,
and absorption line strengths, to be discussed in a future paper
(\cite{kelson99c}). In the future, larger surveys will provide a more
coherent picture in which several hundreds of such galaxies can be
analyzed simultaneously, leading to precision measurements of the
evolving mass function, constraining the detailed evolution of $M/L$
ratios and merger rates in both the field and in clusters.

\acknowledgements

We gratefully acknowledge the comments of the anonymous referee who
helped improve the presentation of this work. We also appreciate the
effort of those at the W.M.Keck observatory who developed and
supported the facility and the instruments that made this program
possible. Support from STScI grants GO05989.01-94A, GO05991.01-94A,
and AR05798.01-94A and NSF grant AST-9529098 is gratefully
acknowledged.

%%%%%%%%%%%%%%%%%%%%%%%%%%%%%%%%%%%%%%%%%%%%%%%%%%%%%%%%%%%%%%%%%%%%%%%%

%%%%%%%%%%%%%%%%%%%%%%%%%%%%%%%%%%%%%%%%%%%%%%%%%%%%%%%%%%%%
\newpage
%\section*{Figure Captions}

\clearpage

\begin{center}
\begin{minipage}{\textwidth}
\vspace{1.3in}
\mkfigbox{5in}{fig2b.ps}
\vspace{0.15in}
\Fig
\end{minipage}
\end{center}
\clearpage

\begin{center}
\begin{minipage}{\textwidth}
\vspace{1.3in}
\mkfigbox{5in}{fig2c.ps}
\vspace{0.15in}
\Fig
\end{minipage}
\end{center}
\clearpage

\begin{center}
\begin{minipage}{\textwidth}
\vspace{1.3in}
\mkfigbox{5in}{fig2d.ps}
\vspace{0.15in}
\Fig
\end{minipage}
\end{center}
\clearpage

\begin{center}
\begin{minipage}{\textwidth}
\vspace{1.3in}
\mkfigbox{5in}{fig2e.ps}
\vspace{0.15in}
\Fig
\end{minipage}
\end{center}
\clearpage

\begin{center}
\begin{minipage}{\textwidth}
\vspace{1.3in}
\mkfigbox{5in}{fig2f.ps}
\vspace{0.15in}
\Fig
\end{minipage}
\end{center}
\clearpage

\clearpage

\clearpage

\begin{center}
\begin{minipage}{\textwidth}
\vspace{1.3in}
\mkfigbox{5in}{fig9b.ps}
\vspace{0.15in}
\Fig
\end{minipage}
\end{center}
\clearpage

\begin{center}
\begin{minipage}{\textwidth}
\vspace{1.3in}
\mkfigbox{5in}{fig9c.ps}
\vspace{0.15in}
\Fig
\end{minipage}
\end{center}
\clearpage

\clearpage

%%%%%%%%%%%%%%%%%%%%%%%%%%%%%%%%%%%%%%%%%%%%%%%%%%%%%%%%%%%%%%%%%%%%%%%%
\clearpage

\begin{deluxetable}{l c c}
\tablewidth{0pt}
\tablecaption{Stellar Templates\label{templates}}
\tablehead{
\colhead{ID} &
\colhead{Spectral} &
\colhead{Date of}\nl
\colhead{} &
\colhead{Type} &
\colhead{Observation}}
\startdata
The Sun (twilight)&G2V&5/96\nl
HD5268&G5IV&8/96\nl
HD102494&G9IV&5/96\nl
HD6833&G9III&8/96\nl
HD72324$^a$&G9III&5/96\nl
HD56224&K1III&5/96\nl
HD4388&K3III&8/96\nl
\enddata
\tablecomments{
Best-fit template
}
\end{deluxetable}

\begin{deluxetable}{l l c l l}
\tablewidth{5in}
\tablecaption{Features used to Measure the Instrumental Resolution
for the Galaxy Spectra
\label{linetable}}
\tablehead{
\colhead{$\lambda$} &
\colhead{Species}&
\colhead{}&
\colhead{$\lambda$} &
\colhead{Species}}
\startdata
5224.137&9-2 P1(2.5)&&6533.044&6-1 P1(2.5) \nl
5238.747&9-2 P1(3.5)&&6553.617&6-1 P1(3.5) \nl
5577.338&[O I]      &&6577.183&6-1 P1(4.5)\nl 
5589.128&7-1 P1(2.5)&&6596.643&6-1 P2(4.5) \nl
5605.143&7-1 P1(3.5)&&6603.990&6-1 P1(5.5)\nl 
5915.301&8-2 P1(2.5)&&6889.288&7-2 P2(1.5) \nl
5924.723&8-2 P2(2.5)&&6900.833&7-2 P1(2.5) \nl
5932.862&8-2 P1(3.5)&&6912.623&7-2 P2(2.5) \nl
5953.420&8-2 P1(4.5)&&6923.220&7-2 P1(3.5) \nl
5970.278&8-2 P2(4.5)&&6939.521&7-2 P2(3.5) \nl
6192.937&5-0 P2(1.5)&&6948.936&7-2 P1(4.5)\nl 
6202.720&5-0 P1(2.5)&&6969.930&7-2 P2(4.5) \nl
6221.771&5-0 P1(3.5)&&6978.258&7-2 P1(5.5)\nl 
6287.434&9-3 P1(2.5)&&7003.858&7-2 P2(5.5) \nl
6300.304&[O I]      &&7011.202&7-2 P1(6.5)\nl 
6363.780&[O I]\nl
\enddata
\tablecomments{
Wavelengths and identifications taken from Osterbrock \etal\ (1997).}
\end{deluxetable}

\begin{deluxetable}{l l c c r r c r r}
\footnotesize
%\small
\tablewidth{0pt}
\tablecaption{Final Velocity Dispersions\label{sigfinal}}
\tablehead{
&&&&&&&\multicolumn{2}{c}{Adopted}\nl
\colhead{ID} &
\colhead{Morphology} &
\colhead{$S/N$ per \AA} &&
\multicolumn{2}{c}{$\sigma_{\rm DF,F}$$^b$ (km/s)} &&
\multicolumn{2}{c}{$\sigma_{\rm DF,K}$$^b$ (km/s)}
}
\startdata
\ \ \,95 & S0   &  48 &&  219.1 &$\pm6.4$ && 220.5 &$\pm6.4$ \nl
     110 & S0/a &  71 &&  171.8 &  3.7 && 174.8 &  3.7 \nl
     129 & S0/a &  65 &&  167.0 &  3.6 && 167.5 &  3.8 \nl
     135 & S0   &  40 &&  161.2 &  5.4 && 159.7 &  5.1 \nl
     142 & S0/a &  43 &&  148.4 &  4.5 && 147.9 &  4.4 \nl
     164 & S0/a &  51 &&  181.0 &  4.5 && 180.6 &  4.3 \nl
     182 & S0   &  28 &&  132.5 &  5.6 && 130.1 &  5.4 \nl
 209$^a$ & Sa   &  35 &&   66.3 &  7.7 &&  73.5 &  7.5 \nl
 209$^a$ & Sa   &  63 &&  109.0 &  5.8 && 107.9 &  5.2 \nl
     211 & S0   &  44 &&  189.1 &  6.2 && 190.5 &  6.4 \nl
     212 & E    &  53 &&  182.3 &  4.9 && 181.0 &  5.7 \nl
     215 & S0   &  33 &&  162.7 &  6.1 && 166.2 &  5.9 \nl
     233 & E/S0 &  74 &&  238.0 &  4.3 && 234.9 &  4.5 \nl
     234 & Sb   &  42 &&  113.1 &  8.5 && 108.9 &  8.2 \nl
     236 & S0   &  43 &&  197.9 &  5.7 && 196.7 &  5.6 \nl
     242 & E    &  58 &&  207.6 &  5.2 && 212.4 &  5.3 \nl
     256 & E    &  93 &&  257.8 &  4.1 && 261.8 &  4.3 \nl
     269 & E/S0 &  92 &&  317.2 &  5.4 && 319.5 &  5.8 \nl
     283 & Un   &  34 &&  147.2 &  5.2 && 145.7 &  5.2 \nl
     292 & S0   &  56 &&  111.3 &  3.5 && 112.3 &  3.2 \nl
     298 & S0   &  85 &&  296.1 &  5.5 && 296.6 &  5.2 \nl
     300 & S0   &  60 &&  254.4 &  5.7 && 248.2 &  5.5 \nl
     303 & E    &  52 &&  164.5 &  4.0 && 163.5 &  4.3 \nl
     309 & E/S0 &  61 &&  229.9 &  4.7 && 228.7 &  4.7 \nl
     328 & Sa   &  59 &&   75.8 &  4.7 &&  81.2 &  4.2 \nl
     335 & S0/a &  37 &&  128.0 &  4.6 && 129.1 &  4.3 \nl
     343 & S0   &  36 &&   63.1 &  5.6 &&  65.1 &  5.0 \nl
     353 & E/S0 &  77 &&  223.6 &  4.3 && 228.5 &  4.4 \nl
     356 & Sa   &  64 &&  191.5 &  3.8 && 190.2 &  3.7 \nl
     359 & S0   &  56 &&  198.7 &  4.4 && 200.3 &  4.5 \nl
     360 & E    &   9 &&  173.7 & 17.7 && 177.5 & 19.7 \nl
     366 & S0/a &  43 &&  226.3 &  6.6 && 228.9 &  6.4 \nl
     368 & Sab  &  21 &&  148.2 &  7.4 && 145.6 &  7.5 \nl
     369 & S0/a &  36 &&  129.7 &  6.3 && 128.2 &  6.1 \nl
     371 & Sa   &  59 &&  175.9 &  3.9 && 174.8 &  4.0 \nl
     372 & Sa   &  45 &&  144.7 &  5.0 && 142.3 &  5.0 \nl
     375 & E    &  78 &&  307.0 &  6.9 && 311.3 &  7.0 \nl
     381 & E/S0 &  48 &&  210.6 &  5.6 && 212.4 &  5.5 \nl
     391 & E/S0 &  62 &&  252.9 &  7.1 && 260.1 &  6.3 \nl
     397 & S0/a &  38 &&  135.1 &  5.2 && 135.9 &  5.1 \nl
     408 & S0   &  47 &&  233.3 &  6.2 && 237.1 &  6.2 \nl
     409 & E    &  21 &&  106.9 &  6.9 && 107.3 &  6.7 \nl
     410 & S0   &  32 &&  150.9 &  5.0 && 148.6 &  5.2 \nl
     412 & E    &  41 &&  170.7 &  6.3 && 169.4 &  6.3 \nl
     433 & S0/a &  37 &&  104.4 &  3.7 && 106.6 &  4.0 \nl
     440 & S0/a &  25 &&   61.0 &  4.8 &&  63.7 &  5.0 \nl
     454 & S0/a &  63 &&  149.1 &  3.8 && 149.6 &  3.7 \nl
     463 & S0   &  57 &&  284.6 &  6.4 && 284.4 &  6.4 \nl
     465 & Sa   &  37 &&  156.3 &  4.7 && 156.7 &  5.0 \nl
     481 & S0   &  25 &&  106.8 &  6.0 && 107.2 &  5.5 \nl
 493$^a$ & E/S0 &  18 &&  118.7 &  9.0 && 116.5 &  8.3 \nl
 493$^a$ & E/S0 &  17 &&  125.1 & 10.1 && 127.5 &  8.8 \nl
     523 & S0/a &  42 &&  171.8 &  5.9 && 174.6 &  5.9 \nl
     531 & E    &  90 &&  285.3 &  4.8 && 281.3 &  5.0 \nl
     534 & E    &  30 &&  120.3 &  6.2 && 120.9 &  6.1 \nl
     536 & E    &  72 &&  252.2 &  5.0 && 254.1 &  5.0 \nl
     549 & Sab  &  30 &&   79.9 &  4.7 &&  79.9 &  4.6 \nl
\enddata
\tablecomments{
Notes:
$\sigma_{\rm DF,F}$ refers to measurements made using an analog of
the Fourier Fitting method except that the $\chi^2$ summation is
performed in the real domain.
$\sigma_{\rm DF,K}$ refers to measurements made using the direct
fitting method of \S \ref{direct}.
(a) Galaxies \#209 and \#493 were observed in two different slit-masks and
position angles. The first listed value for \#209 was derived from a
mis-aligned slitlet (see \S \ref{consist}).
(b) These values have been corrected for aperture size to a nominal
aperture with diameter of 3\Sec 4 at the distance of Coma.
}
\end{deluxetable}

\end{document}